\documentclass[english,superscriptaddress,twocolumn]{revtex4}
\usepackage{mathptmx}

\usepackage[T1]{fontenc}
\usepackage[latin9]{inputenc}
\usepackage{amsmath}
\usepackage{graphicx}
\usepackage{amssymb}
\usepackage{esint}
\usepackage{color}
\makeatletter

\DeclareFontEncoding{LGR}{}{}

\newcommand{\lyxmathsym}[1]{\ifmmode\begingroup\def\b@ld{bold}
  \text{\ifx\math@version\b@ld\bfseries\fi#1}\endgroup\else#1\fi}

\@ifundefined{textcolor}{}
{%
 \definecolor{BLACK}{gray}{0}
 \definecolor{WHITE}{gray}{1}
 \definecolor{RED}{rgb}{1,0,0}
 \definecolor{GREEN}{rgb}{0,1,0}
 \definecolor{BLUE}{rgb}{0,0,1}
 \definecolor{CYAN}{cmyk}{1,0,0,0}
 \definecolor{MAGENTA}{cmyk}{0,1,0,0}
 \definecolor{YELLOW}{cmyk}{0,0,1,0}
 }

\makeatother

\usepackage{babel}

\begin{document}

\title{Charm quarks in medium and their contribution to di-electron spectra
in relativistic heavy ion collisions}

\author{Hao-jie Xu}
\affiliation{Department of Modern Physics, University of Science and Technology
of China, Anhui 230026, People's Republic of China}

\author{Xin Dong}
\affiliation{Nuclear Science Division, Lawrence Berkeley National Laboratory,
Berkeley, California 94720, USA}

\author{Li-juan Ruan}
\affiliation{Physics Department, Brookhaven National Laboratory, Upton, New York
11973-5000, USA}

\author{Qun Wang}
\affiliation{Department of Modern Physics, University of Science and Technology
of China, Anhui 230026, People's Republic of China}

\author{Zhang-bu Xu}
\affiliation{Physics Department, Brookhaven National Laboratory, Upton, New York
11973-5000, USA}

\author{Yi-fei Zhang}
\affiliation{Department of Modern Physics, University of Science and Technology
of China, Anhui 230026, People's Republic of China}

\begin{abstract}
We study the dynamics of charm quarks in the partonic medium and its
implication to the di-electron spectra in high energy heavy ion collisions.
The charm quarks traversing a thermal medium is simulated by the
relativistic Langevin equation for elastic scatterings of charm quarks
by thermal partons in an expanding fireball. The transport
coefficients of charm quarks are calculated by the in-medium T-matrix
method, where a static heavy quark potential is used with parameters
fitted by the lattice QCD results. The di-electron invariant mass
spectra are computed in most central collisions and are compared
to the STAR data. The angular correlations of di-electrons
are almost the same in $p+p$ and Au+Au collisions in the mass range $1.1<M<2.5~\mathrm{GeV/c^2}$
with the back-to-back feature. This means that the angular correlation is
intact even with medium interaction at the RHIC energy.
\end{abstract}
\maketitle

\section{Introduction}

The goal of heavy ion collision experiments at Relativistic Heavy
Ion Collider (RHIC) \cite{Adams:2005dq,Adcox:2004mh} and Large Hadron
Collider (LHC) \cite{Evans:2008zzb} is to search for and then study
properties of the new state of matter, the Quark-Gluon Plasma (QGP).
The jet quenching and strong elliptic flow observed at RHIC \cite{Ackermann2001,Adcox2002}
and LHC \cite{Aamodt:2010pa} indicate that the hot and dense medium
interacts strongly and behaves as a nearly prefect fluid \cite{Gyulassy:2004zy,Shuryak:2008eq},
so it is called the strongly coupled QGP or sQGP. Among all observables
to pin down the sQGP, electromagnetic probes such as photons and dileptons
are expected to provide clean signatures due to their weak couplings
to the hot and dense matter \cite{McLerran1985a,Kajantie1986dh}.

The dilepton invariant mass spectrum is usually divided into the low,
intermediate and high mass regions (LMR, IMR and HMR), based on the
notion that each region is dominated by different sources. In the
LMR, $M\lesssim 1~\mathrm{GeV/c^2}$, the medium modification of the $\rho$ meson
spectral function is the key to describe the di-muon enhancement in
the NA60 experiment at the Super Proton Synchrotron (SPS) \cite{vanHees:2006ng,Ruppert2008,Dusling2007},
as well as the di-electron enhancement in the STAR experiment at RHIC
\cite{Xu2012,Linnyk:2011vx}. In the IMR, $1\lesssim M\lesssim3~\mathrm{GeV/c^2}$, the thermal quark-antiquark annihilation in the QGP phase was
proposed to provide a measurable signal for the de-confinement phase
transition at the RHIC energy \cite{Deng2011}. However, in this mass
region, the dilepton yields from semi-leptonic decays of open charm
hadrons increase rapidly with collisional energies. In Ref.~\cite{Xu2012}
by some of us, a naive model was used to estimate electrons from open
charm decays. It was found that electrons from open charm hadron
decays out-populate the thermal ones at the RHIC energy in the IMR.

The charm and bottom quarks are  not expected to fully equilibrate
in the hot and dense medium, so they are regarded as hard probes to
the partonic medium due to their large mass scales compared to the
temperature. Perturbative Quantum Chromodynamics (PQCD) calculations
predicted a less energy loss for heavy quarks than for light quarks
\cite{Wicks2007} due to the {}``dead-cone'' effect \cite{Dokshitzer2001}.
But the measurements of non-photonic electrons from semi-leptonic
decays of heavy flavor hadrons at RHIC \cite{Adare2007} give strongly
suppressed nuclear modification factor $R_{AA}$ and large elliptic
flow $v_{2}$. This implies a substantial modification of heavy quark
spectra when they traverse the hot and dense medium.

There are a variety of models on the market for heavy quarks in partonic
medium, such as the heavy quark diffusion model using the Fokker-Planck-Langevin
equation \cite{Alberico2011c,Hees2005,Hees2008,He2012} and the model
based on the Boltzmann transport simulation \cite{Uphoff2010,Uphoff2012},
etc.. These models show that in order to obtain the same suppression
of $R_{AA}$ for heavy quarks as for light quarks, much larger transport
coefficients of heavy quarks than standard PQCD prediction have to
be used in the diffusion process. This implies non-perturbative effects.
The in-medium T-matrix method is one of non-perturbative models for
heavy-light and heavy-gluon interactions \cite{Mannarelli2005,Hees2008,Riek2010,Huggins2012}
with a static heavy quark potential extracted from the lattice QCD
(LQCD) with relativistic corrections. This reduces the thermalization
time of heavy quarks 3-4 times shorter than the PQCD prediction.

In high energy heavy ion collisions, charm quark pairs are produced
back to back in their center of mass frame in the gluon fusion process,
$gg\rightarrow c\bar{c}$. The angular correlation of charm quark
pairs are expected to be modified by the interaction of charm quarks
with the surrounding partonic medium \cite{Zhu2007,Zhu2008}. In the
previous work by some of us \cite{Xu2012}, the medium modification
of the angular correlation was neglected. In this paper, we will adopt
a more realistic description for the dynamics of charm quarks in partonic
medium and then provide a better model for the  di-electrons
from open charm hadron decays.

This paper is organized as follows. In Sec.~\ref{sec:dilepton}, we
give an introduction about the thermal dilepton production with the charm background
in high energy heavy ion collisions. In Sec.~\ref{sec:charm}, we give an introduction of the relativistic Fokker-Planck-Langevin equation for
charm quark diffusion in the partonic medium. The space-time evolution can be described by
a (2+1)-dimension hydrodynamical model. In Sec.~\ref{sec:observable} we calculate observables
related to open charm hadrons, such as transverse momentum spectra of $D^{0}$
mesons and the nuclear modification factor for electrons from semi-leptonic
decays of open charm hadrons. We calculate the invariant mass spectra
of di-electrons including the charm background with medium effects. We finally give a summary
of our results in Sec.~\ref{sec:Summary}.

\section{Thermal dilepton production with charm background}

\label{sec:dilepton}
In this section we give an introduction about
the thermal dilepton production with the charm background
in high energy heavy ion collisions. The production rate of
thermal dilepton in heavy ion collisions is given by  \cite{Xu2012,Deng2011}
\begin{eqnarray}
\frac{dN_{ll}}{d^{4}xd^{4}p}&=&-\frac{\alpha}{4\pi^{4}}\frac{1}{M^{2}}n_{B}(p\cdot u)\left(1+\frac{2m_{l}^{2}}{M^{2}}\right)\nonumber\\
&&\times\sqrt{1-\frac{4m_{l}^{2}}{M^{2}}}\mathrm{Im}\Pi^{R}\left(p,T\right).
\end{eqnarray}
where $m_{l}$ is the lepton mass, $\alpha =1/137$ is the electromagnetic
fine structure constant, $M$ and $p$ are the dilepton invariant
mass and four momentum vector respectively, $\Pi^{R}$ is the retarded polarization
tensor from the quark loop in the partonic phase or hadronic loops in the hadronic
phase. For the hadronic phase, we include the medium modifications
of vector mesons from scatterings of vector mesons by hadrons in the
thermal medium as was done in Ref. \cite{Xu2012}.
The Bose-Einstein distribution function $n_{B}(p\cdot u)=1/(\exp(p\cdot u/T)-1)$ depends on
the fluid velocity $u$ and the local temperature $T$ of the thermal medium
as functions of space-time, which are described by a (2+1)-dimension ideal
hydrodynamical model \cite{Deng2011,Xu2012}.
The distribution of the initial energy density is determined by the
Glauber model with $5\%$ of the contribution from binary collisions.
We use the LQCD equation of state (EOS), namely S95P-PCE, which
has a wide range of phase transition temperatures from 184 to 220
MeV. The chemical and kinetic freeze-out temperatures are set to
$T_{\mathrm{chem}}=150~\mathrm{MeV}$ and $T_{\mathrm{f}}=106~ \mathrm{MeV}$.
We focus on most central collisions of Au+Au at $\sqrt{s}=200$ GeV and set
the impact parameter as $b=2.4$ fm. The initial energy density at the center of the overlapping region at $b=0$ is set to $\epsilon_{0}=55~\mathrm{GeV/fm^{3}}$
and the equilibrium time to $\tau_{0}=0.4$ fm
by fitting the transverse momentum spectra of long-lived hadrons \cite{Xu2012}.

\begin{figure}
\begin{centering}
\includegraphics[scale=0.45]{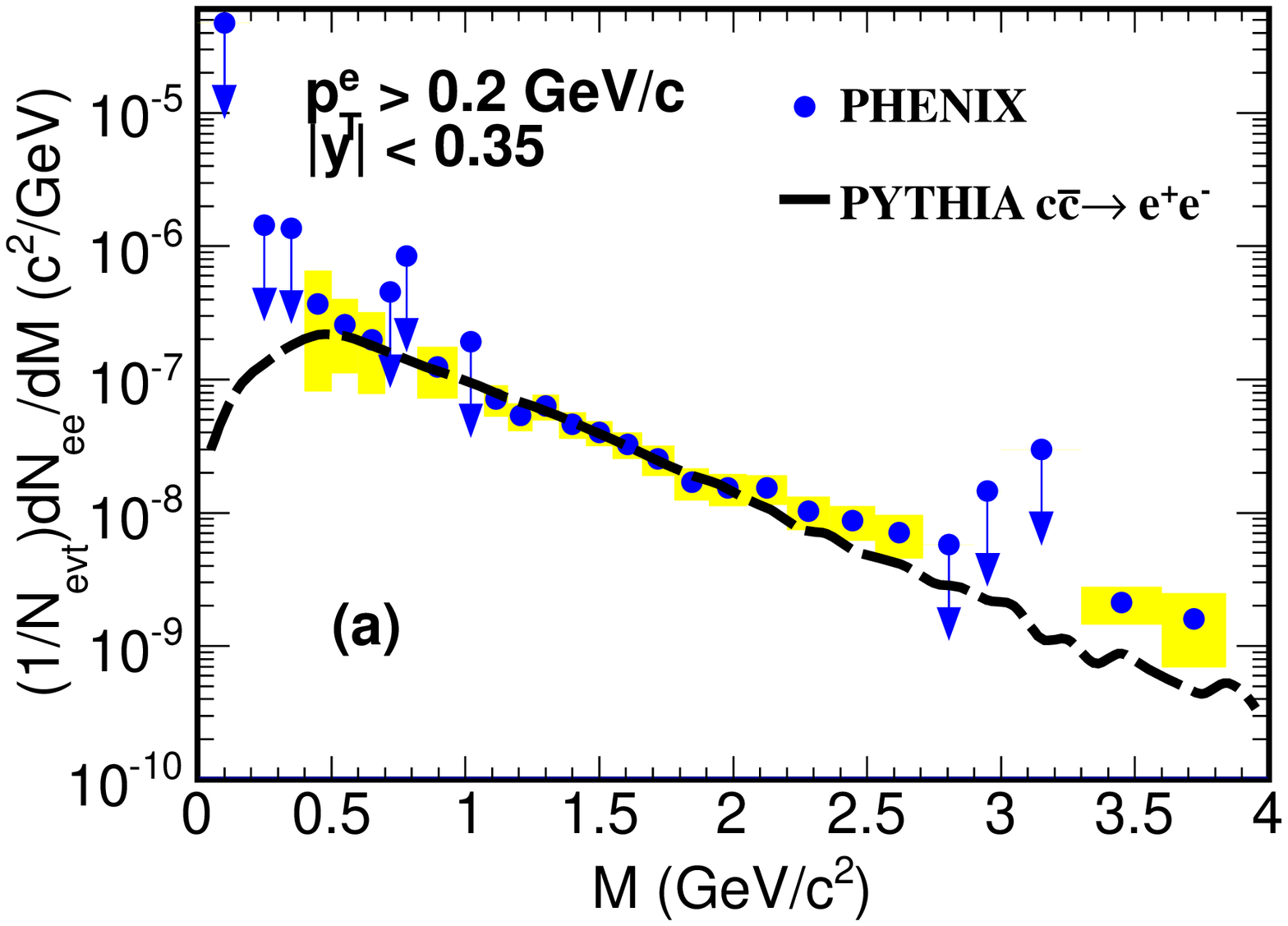}
\includegraphics[scale=0.45]{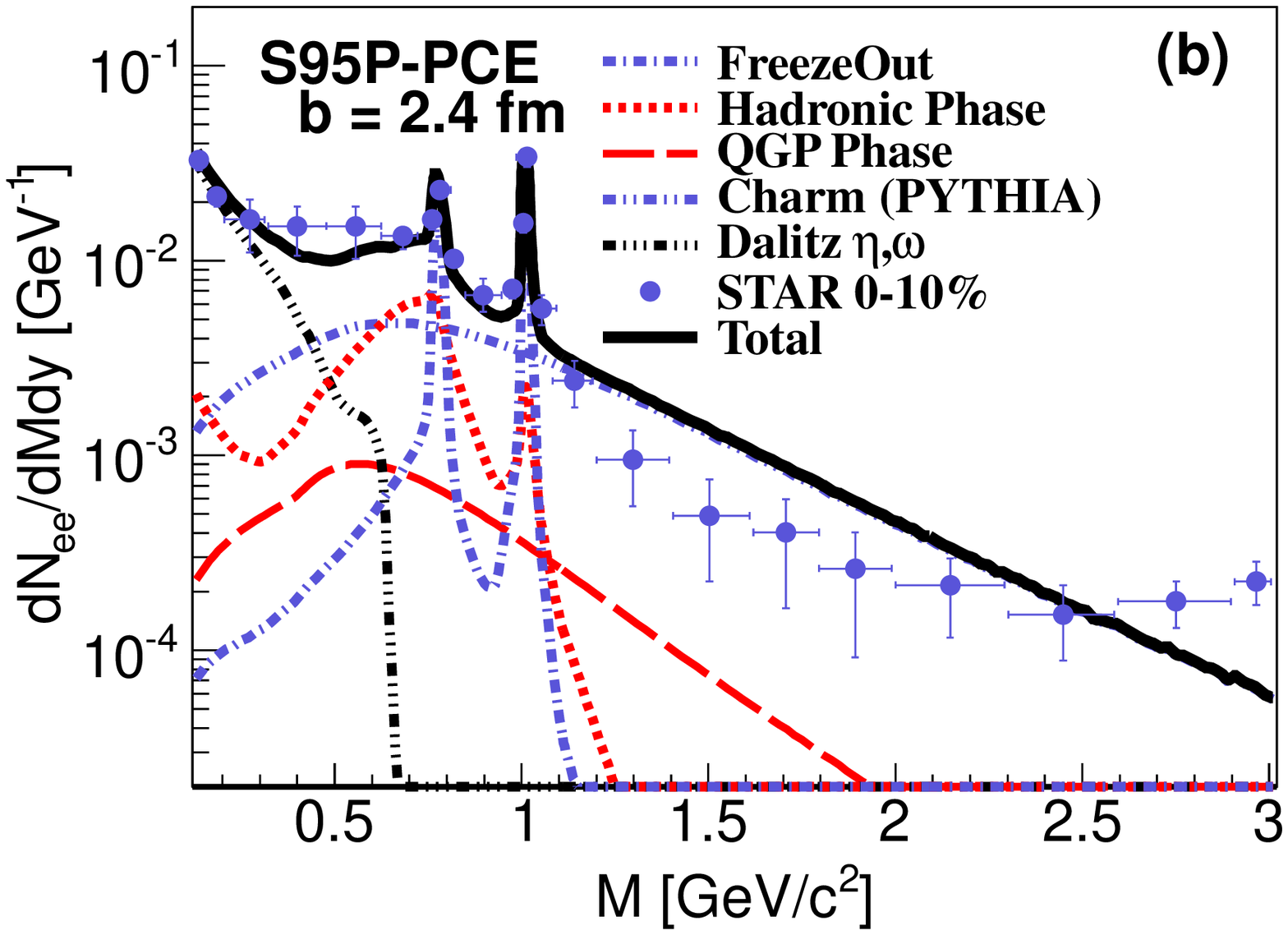}
\par\end{centering}
\caption{(Color online) (a) The di-electron cross section from semi-leptonic decays of
open charm hadrons in p+p collisions by PYTHIA with the PHENIX detector acceptance.
The data are taken from PHENIX \cite{Adare2009}. (b) The invariant mass spectra and
comparison with the STAR data \cite{Zhao2011,Adamczyk:2013caa} in most central Au+Au collisions
with the STAR acceptance.
\label{fig:ppcollision}}
\end{figure}

The semi-leptonic decays of open charm hadrons contribute to the dilepton background.
Such a background cannot be removed by current experiments.
To simulate this background in our calculations, we use the Monte Carlo
event generator PYTHIA \cite{Sjostrand:2000wi}~(set MSEL=4, i.e. choose the
charm production with massive matrix elements) for p+p collisions.
The dilepton yield in the mass range $[1.1,2.5]~\mathrm{GeV/c^{2}}$ is
dominated by semi-leptonic decays of open charm hadrons. In the PHENIX
acceptance the integrated yield of di-electrons per event from heavy-flavor
decays in that range is $(4.21\pm 0.28\pm1.02)\times10^{-8}$ \cite{Adare2009}.
With the branching ratio of charm quarks to electrons \cite{Eidelman2004}
and the correction to the geometrical acceptance, the rapidity density
of $c\bar{c}$ pairs can be estimated \cite{Adare2009}. We use the
PYTHIA event generator with the PHENIX acceptance to reproduce
in Fig.~\ref{fig:ppcollision}(a) the di-electron spectra from open charm hadrons in p+p collisions. The result for heavy ion collisions is obtained by multiplying the p+p collision result by the binary collision number of heavy ion collisions as shown in Fig. Fig.~\ref{fig:ppcollision}(b).

We compute the di-electron spectra in most central collisions with STAR
acceptance: $p_{T}>0.2~ \mathrm{GeV/c}$,  $|y^{ee}|<1$ and $|\eta^{e}<1|$ for
the transverse momentum and rapidity of di-electrons and the pseudo-rapidity of individual electrons, respectively, see Fig.~\ref{fig:ppcollision}(b).
We include vector mesons decays after thermal freeze-out (blue-dash-dotted line) and  the Dalitz decays of $\eta$ and $\omega$ (black-dash-dot-dot-dotted line).
We do not include the medium modifications of charm quarks here. By contrast
a simple parametrization was used in Ref. \cite{Xu2012}
about the nuclear modification factor for single electrons from open charm decays.
As shown in the figure, the charm background (blue-dash-dot-dotted line) dominates
over the contributions from the QGP phase (red-long-dashed line) and the hadronic phase (red-dashed line). The $p_{T}$ spectra of charm quarks as well as the angular correlation of the $c\bar{c}$ pairs can be modified by their interaction with the thermalized light partons. This has to be described in a more realistic and reasonable model.
This is what we will do in this paper.

\section{Charm quarks in partonic medium}
\label{sec:charm}
In this section we will describe our model for charm quarks traversing the partonic medium
based on the Langevin equation for heavy quark diffusion and the (2+1)-dimension hydrodynamical
model for the fireball expansion.

\subsection{Fokker-Planck-Langevin equation }

The charm quarks traversing in a partonic medium will change their
momentum distribution by interaction with thermal partons. Due to
its heavy mass, the movement of a charm quark in the partonic medium
can be treated as Brownian motion governed by the Langevin equation
\cite{Svetitsky1988,Dunkel2009a},
 \begin{eqnarray}
dx^{i} & = & \frac{p^{i}}{E}dt,\label{eq:Langevinx}\\
dp^{i} & = & -a_{j}^{i}p^{j}dt+c_{j}^{i}\odot dB^{j}\left(t\right),\label{eq:Langevinp}\end{eqnarray}
where $x^{i}$ and $p^{i}$ denote the i-th spatial components of
the position and momentum vectors $\mathbf{x}$ and $\mathbf{p}$
of the charm quark, $a_{j}^{i}$ and $c_{j}^{i}$ are coefficients
which are related to drag force and diffusion, $dB^{i}\left(t\right)$
are noise variables which are specified by their correlations
 \begin{eqnarray}
\left\langle dB^{i}(t)\right\rangle  & = & 0,\nonumber \\
\left\langle dB^{i}(t)dB^{j}(t')\right\rangle  & = & \begin{cases}
\delta_{ij}\sqrt{dt}, & t=t'\\
0, & t\neq t'\end{cases},\end{eqnarray}
We can choose the standard Gaussian noise satisfying the above correlation,
\begin{equation}
\mathcal{P}\left(dB(t)\in[y,y+dy]\right)=\left(\frac{1}{2\pi dt}\right)^{3/2}\exp\left(-\frac{y^{2}}{2dt}\right),\end{equation}
where $dB(t)=\sqrt{\sum_{i}(dB^{i})^{2}}$. The symbol $\odot$ in
Eq. (\ref{eq:Langevinp}) denotes the discretization rule.

We use post-point discretization, then the corresponding Fokker-Planck
equation is \cite{Dunkel2009a}
\begin{equation}
\frac{\partial}{\partial t}f=\frac{\partial}{\partial p^{i}}\left[\left(a_{j}^{i}p^{j}-c_{r}^{k}\frac{\partial}{\partial p^{k}}c_{r}^{i}\right)f+\frac{1}{2}\frac{\partial}{\partial p^{k}}\left(c_{r}^{k}c_{r}^{i}f\right)\right]\label{eq:LangBrown}\end{equation}
where $f$ is the phase-space distribution of heavy quarks. If the
heat bath is stationary, isotropic and homogeneous, the coefficients
take the simple diagonal form \cite{Dunkel2009a}
\begin{equation}
a_{j}^{i}=\Gamma(E)\delta_{j}^{i},\ c_{i}^{j}=\sqrt{2D(E)}\delta_{j}^{i},\label{eq:a-c-simple}\end{equation}
where $\Gamma(E)$ and $D(E)$ are the drag and diffusion coefficient
respectively. The generalized fluctuation-dissipation relation (the
equilibrium condition) gives
 \begin{equation}
D(E)=ET\Gamma(E),\end{equation}
where $E=\sqrt{\mathbf{p}^{2}+m_{Q}^{2}}$ is the charm quark energy.
We will use the form (\ref{eq:a-c-simple}) for $a_{j}^{i}$ and $c_{j}^{i}$
in our calculation.

The Fokker-Planck equation can be derived from the Boltzmann equation
in the Landau approximation \cite{Svetitsky1988},
 \begin{equation}
\frac{\partial f}{\partial t}=\frac{\partial}{\partial p^{i}}\left[A(E)p^{i}f+\frac{\partial}{\partial p^{j}}\left(B^{ij}(\mathbf{p})f\right)\right],\label{eq:LangBoltz}\end{equation}
where the relaxation rate is given by
 \begin{eqnarray}
A(E) & = & \frac{1}{2E_{p}}\int\frac{d^{3}\mathbf{q}}{\left(2\pi\right)^{3}E_{q}}f_{i}\left(x,q\right)\nonumber \\
 &  & \times\int\frac{d^{3}\mathbf{p}'}{\left(2\pi\right)^{3}E_{p'}}\int\frac{d^{3}\mathbf{q}'}{\left(2\pi\right)^{3}E_{q'}}\left(1-\frac{\mathbf{p}\cdot\mathbf{p}'}{\mathbf{p}^{2}}\right)\nonumber \\
 &  & \times|\mathcal{M}(s)|^{2}(2\pi)^{4}\delta^{4}(p+q-p'-q'),\label{eq:friction}\end{eqnarray}
where $f_{i}\left(x,q\right)$ is the phase-space distribution for
light quarks or gluons, and $q$ and $q'$ denote their 4-momenta.
The heavy quark 4-momenta are denoted by $p,p'$ whose spatial components
are $\mathbf{p},\mathbf{p}'$ respectively. The function $B^{ij}(\mathbf{p})$
is a momentum integral similar to $A(E)$. Comparing with Eqs. (\ref{eq:LangBrown},\ref{eq:LangBoltz}),
we have
 \begin{eqnarray}
\Gamma(E) & = & A(E)+\frac{1}{E}\frac{\partial}{\partial E}D(E),\end{eqnarray}
where we used $\partial/\partial p^{k}=\left(p^{k}/E\right)\partial/\partial E$.
In Eq. (\ref{eq:friction}) $\mathcal{M}\left(s\right)$ is the scattering
amplitude of a heavy quark by a light quark or a gluon. Note that
$A(E)$ and $\Gamma(E)$ depend on temperature via $f_{i}\left(x,q\right)$
and $\mathcal{M}(s)$.

\subsection{Charm quark diffusion in expanding partonic medium}
\subsubsection{Charm quark transport coefficients}
As mentioned in the introduction, the PQCD calculation of
heavy quark transport coefficients cannot reproduce the data for the
heavy quark $R_{AA}$ and the elliptic flow, so non-perturbative effects
have to be included. To this end, we adopt the in-medium T-matrix
method based on the static heavy quark potential \cite{Mannarelli2005,Hees2008}
to calculate $\mathcal{M}\left(s\right)$. The potential is given
by a microscopic model with a Coulomb and a confinement component,
where the free parameters are fixed by comparison with the color-averaged
free energy of charm quarks from LQCD \cite{Riek2010}.

\begin{figure}
\begin{centering}
\includegraphics[scale=0.45]{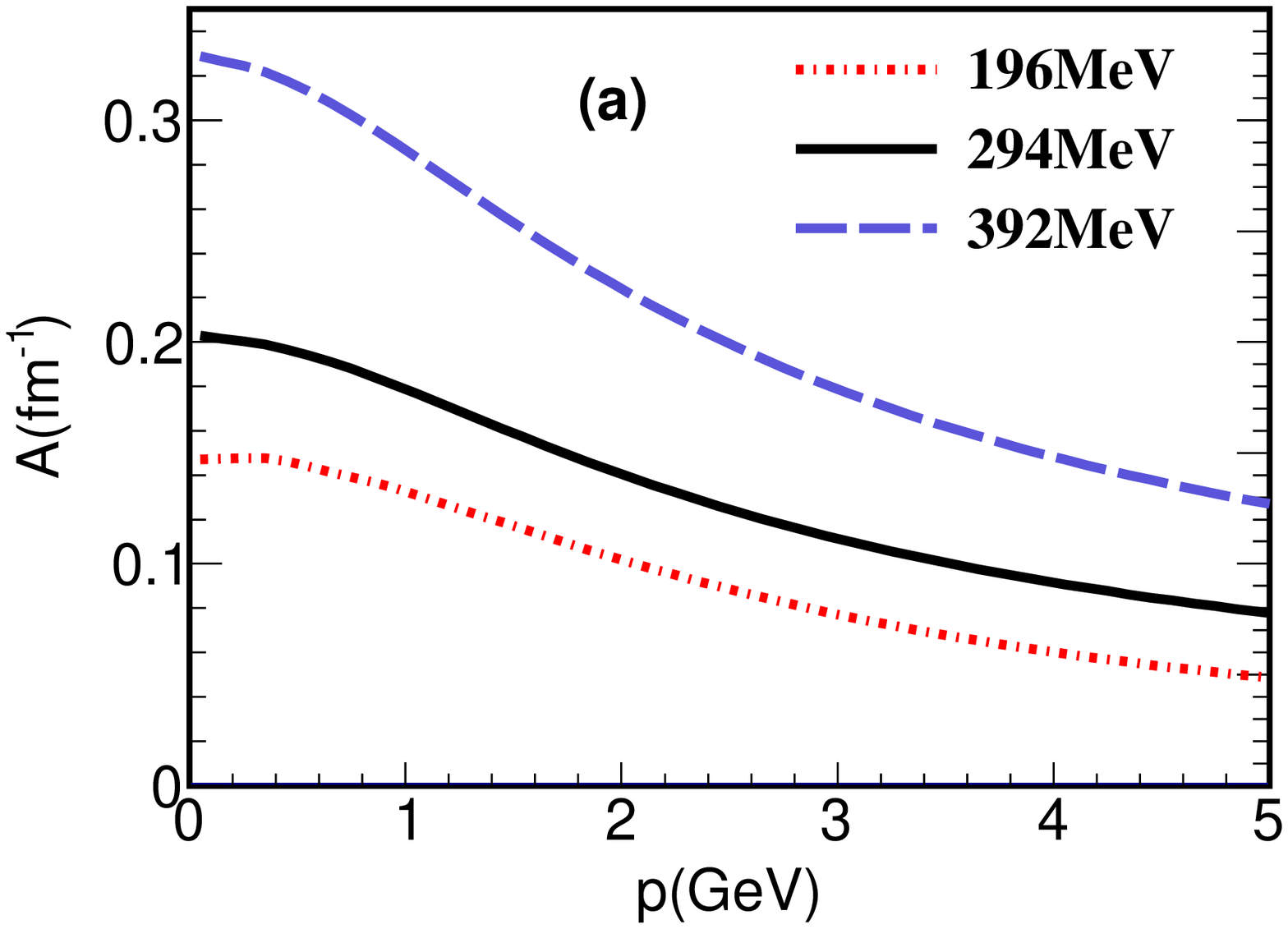}
\par\end{centering}

\begin{centering}
\includegraphics[scale=0.45]{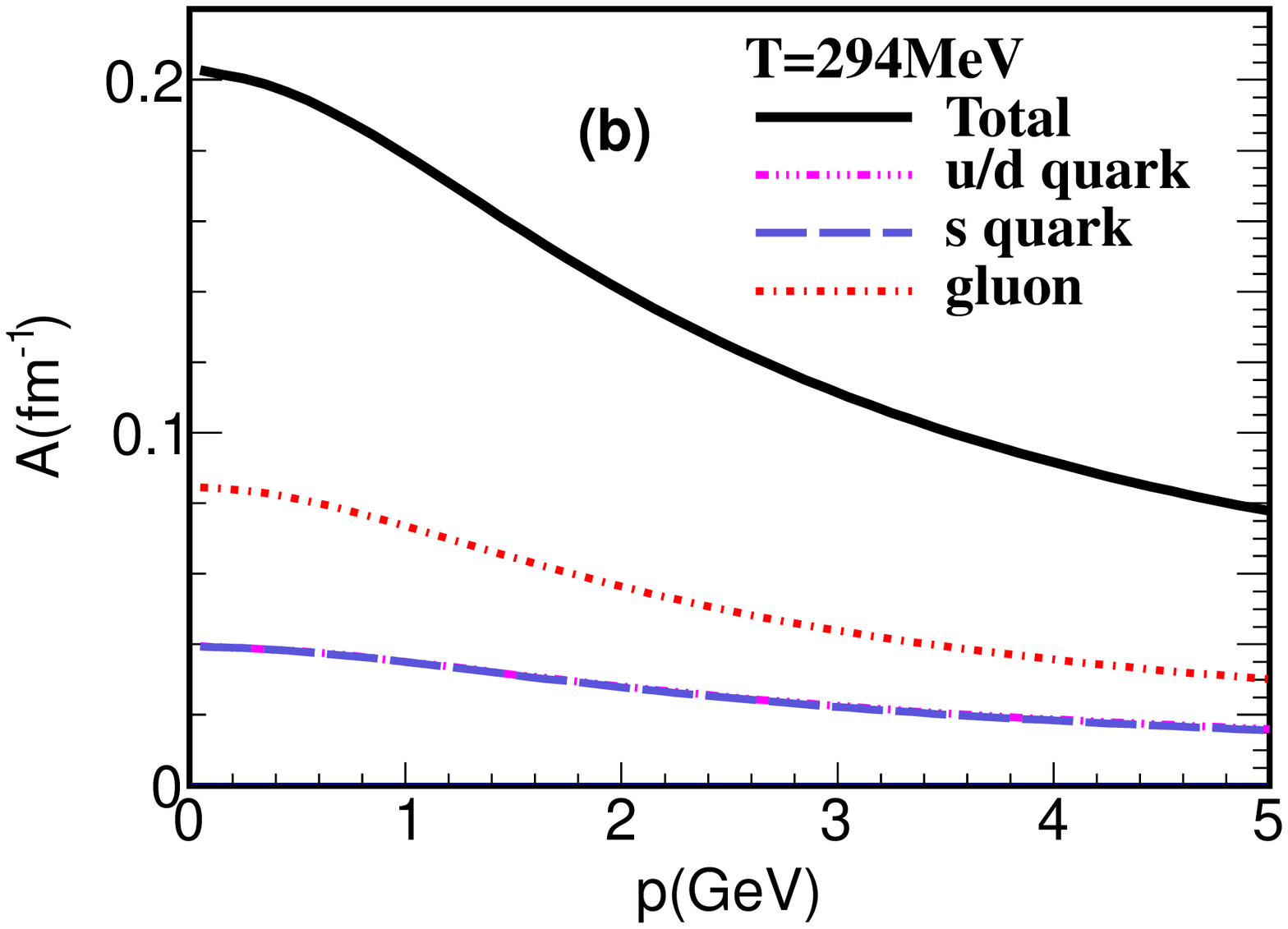}
\par\end{centering}
\caption{(Color online) (a) Charm quark relaxation rates as functions of 3-momenta at different
temperatures. (b) Charm quark relaxation rates from scatterings by
light and strange quarks and gluon. The curves for s quarks and u/d quarks are almost identical and
indistinguishable. The temperature is set to $T=294$~MeV. \label{fig:RelaxationRate}}
\end{figure}

We choose ``potential-1'' in Ref. \cite{Riek2010} which corresponds
to the (2+1)-flavors LQCD result. The internal energy is used as
the interaction kernel in the T-matrix equation. The 4-dimensional
Bethe-Salpeter type T-matrix equation can be reduced to 1-dimensional
equation by the Thompson method and partial-wave expansion. The charm
quark scattering with light quarks \cite{Riek2010} and gluons \cite{Huggins2012}
are all considered. In Fig. \ref{fig:RelaxationRate}(a) we show the
relaxation rate of the charm quark as functions of 3-momenta at different
temperatures. The contributions from $u/d$ and $s$ quarks and from
gluons at $T=294$ MeV are shown in Fig. \ref{fig:RelaxationRate}(b).
The relaxation rate given by the in-medium T-matrix is much larger
than the PQCD result \cite{Huggins2012}. This is because the charm-light
quark scattering amplitude shows a Feshbach resonance structure in
the T-matrix calculation \cite{Riek2010}. The resonance feature greatly
enhances the cross section of heavy-light quark scattering and shortens
the charm quark equilibration time.

\subsubsection{Initial and freeze-out condition}
We use a 2+1 dimension ideal hydrodynamical model to simulate the thermal medium
for both the calculation of dilepton production rate and simulation of charm quarks' movement in medium.
Instead of using PYTHIA simulation as in our previous work summarized in Sec.~\ref{sec:dilepton},
we generate the charm quark pairs by PYTHIA, and we put them to the thermal medium and let them evolve
under Langevin evolution, then we put them back to PYTHIA for them to undergo the following process.
Before the equilibration time $\tau_{0}$, we assume charm quarks move as free streaming.

We choose the freeze-out temperature of open charm hadrons to be the transition temperature
$T_{c}=184~\mathrm{MeV}$. Actually there is no rigorous definition of the
transition temperature for a crossover manifested by the lattice equation
of state. By the transition temperature here we mean the hadronic/partonic
phase below/above it. Note that the freeze-out temperature of light
hadrons ($=106~\mathrm{MeV}$) is much lower than the transition temperature. We assume that
charm quarks hadronize at the freeze-out temperature to form charm
hadrons (mostly open charm mesons) and decouple from the medium immediately.
It is shown in Ref. \cite{Das:2013lra} that if open charm hadrons
undergo further interaction with the hadronic medium, their transverse
momentum spectra are suppressed by about $20-25\%$ for $p_{T}=3-10~\mathrm{GeV/c}$ at RHIC energy.
We will discuss below that such an effect can
be partially achieved by decreasing the freeze-out temperature.

\subsubsection{Hydro-Langevin evolution}
To solve the Fokker-Planck-Langevin equation in a expanding fluid,
at each space-time point, we boost the charm quark momentum to the
local rest frame of the fluid cell at the position of the charm quark,
then let the charm quark undergo Brownian motion and change its position
and momentum, and finally boost it back to the lab frame \cite{He2012}.
So the phase space state of the charm quark is traced in the lab frame.
The time interval for the position and momentum update in the Langevin
equation are kept equal to that for the temperature and fluid velocity
update in hydrodynamical evolution. We set it to $d\tau=0.01$ fm
in the lab frame. The time interval in one fluid cell is given by
\cite{Dunkel2009}
\begin{equation}
\Delta\tau=\gamma d\tau\left(E_{Q}-p_{1}v_{1}-p_{2}v_{2}\right)/E_{Q}
\end{equation}
where $E_{Q}$,$p_{1}$,$p_{2}$ are the energy and the spatial components
of a charm quark momentum, $v_{1}$ and $v_{2}$ are the spatial components
of the fluid velocity at the charm quark position, and
$\gamma=1/\sqrt{1-v_{1}^{2}-v_{2}^{2}}$
is the Lorentz contraction factor. All above quantities are defined
in the lab frame. We then put charm quarks to perform the Hydro-Langevin
simulation step by step. In each step, the momentum diffusion of the
charm quark is controlled by the drag and diffusion coefficients as
functions of the charm quark's 3-momentum and the temperature of the
fluid cell at the position of the charm quark.

As long as the local temperature of the fluid cell at the position
of the charm quark is below $T_{c}$ , we stop the Langevin evolution
of the charm quark and record its final position and momentum. After
all charm quarks complete their evolution, we obtain the $p_{T}$ spectra
as well as the angular correlation of charm quark pairs.

\begin{figure}
\begin{centering}
\includegraphics[scale=0.45]{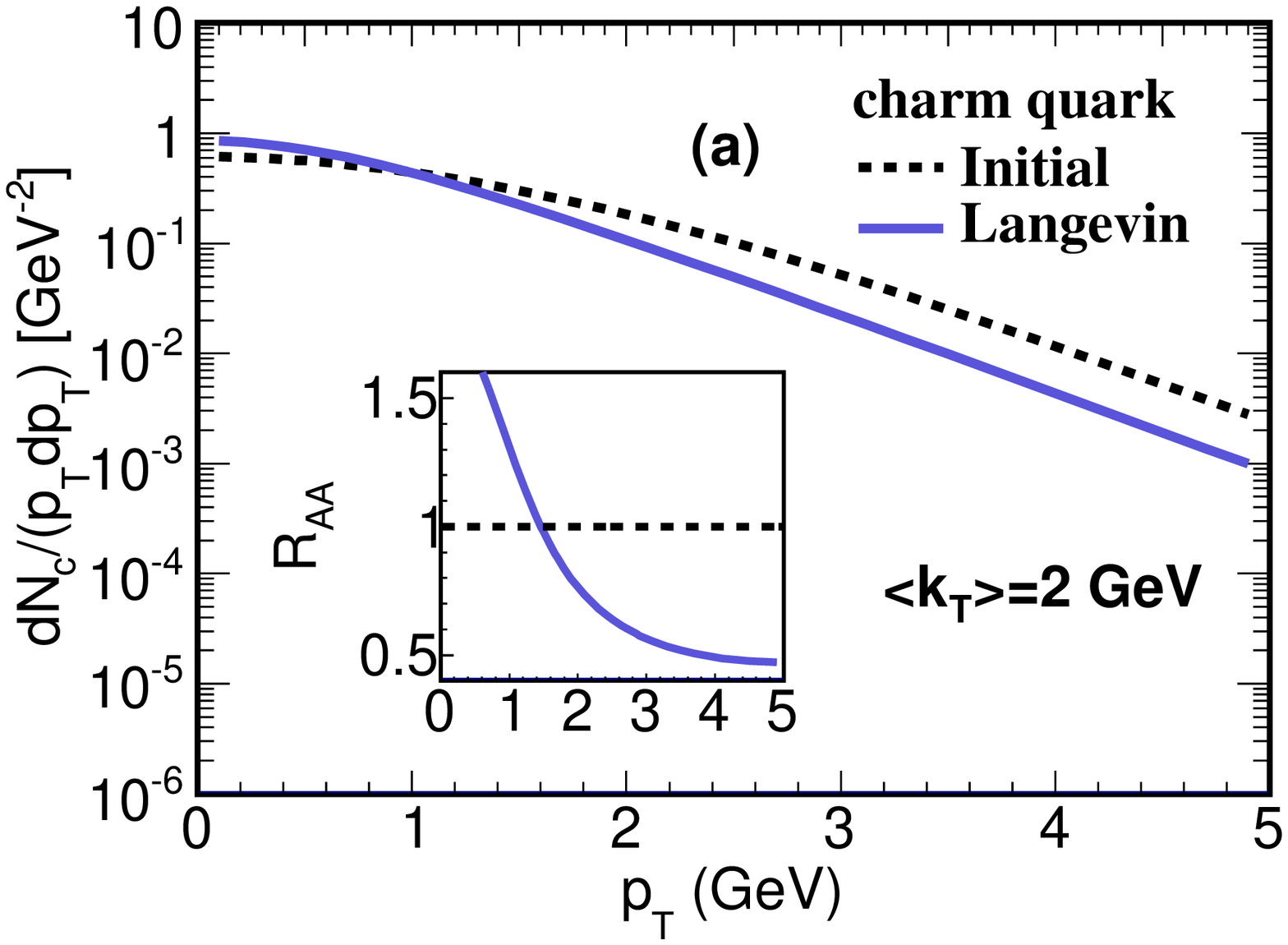}
\par\end{centering}

\begin{centering}
\includegraphics[scale=0.45]{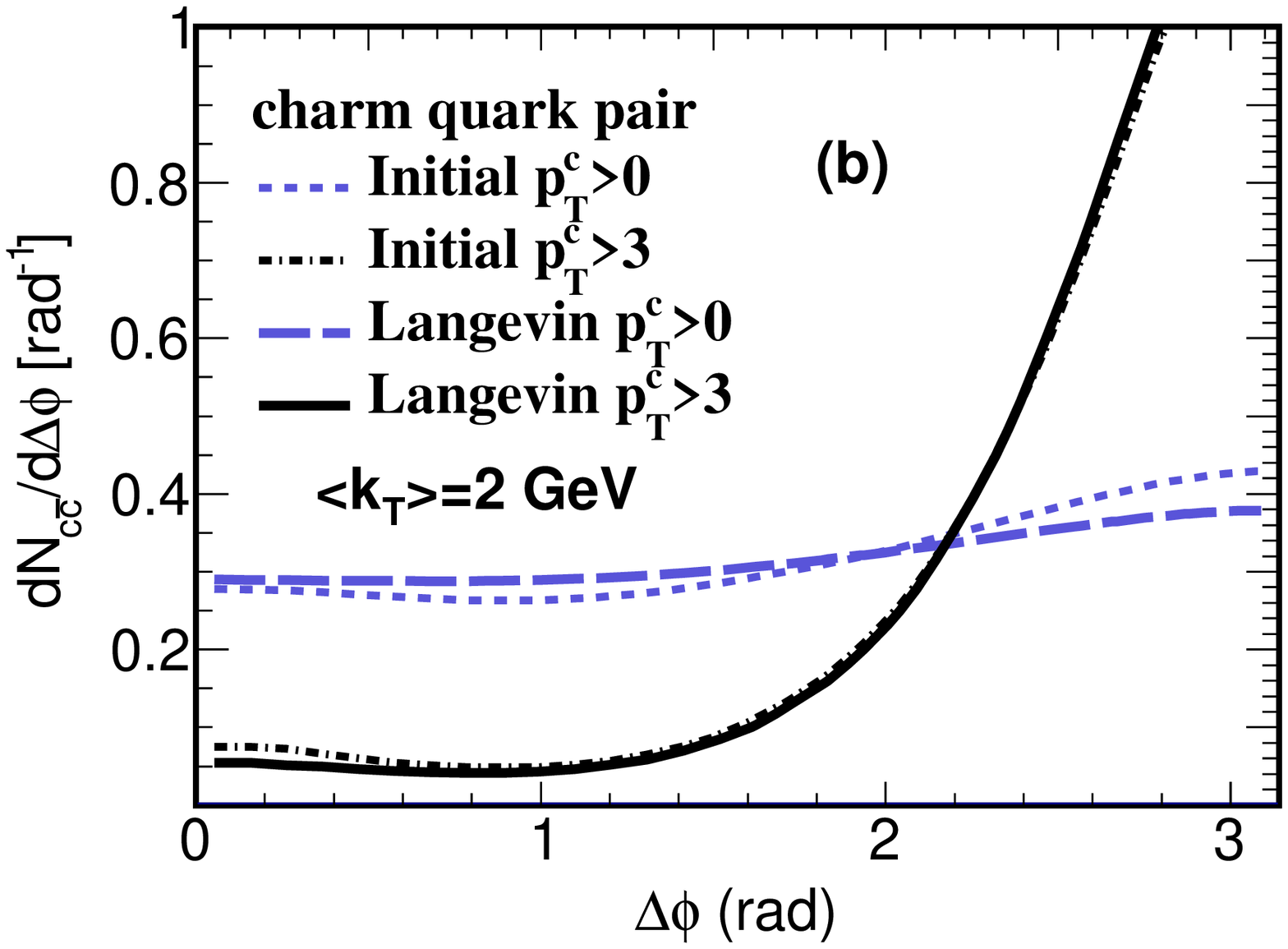}
\par\end{centering}

\caption{(Color online) The $p_T$ spectra and azimuthal angular correlation of charm quarks.
The integrated multiplicities are normalized to unit. (a) The $p_{T}$ spectra of charm quarks in the initial state without
medium modification and those after Hydro-Langevin evolution in the
final state. The nuclear modification factor is shown in the inset.
(b) The azimuthal angular correlation of charm quark pairs in the initial and
final states. The different $p_{T}$ cutoffs are chosen. The freeze-out
temperature is set to $T_{c}=184$ MeV. \label{fig:charmquark}}

\end{figure}

\subsection{Numerical result and parameter dependence}
\subsubsection{ $R_{AA}$ and correlation in azimuthal angles}
In Fig. \ref{fig:charmquark}
we show the final $p_{T}$ distribution of charm quarks and the
correlation in azimuthal angles for charm quark pairs in Hydro-Langevin simulation.
The comparison with the PYTHIA results is also given.
The nuclear modification factor $R_{AA}$, which measures the charm quark energy loss in medium,
is shown in the inset of Fig. \ref{fig:charmquark}(a).

We can see in Fig. \ref{fig:charmquark}(b) that the medium modification
of the correlation in azimuthal angles of charm quark pairs is small in both low
and high $p_{T}$ range. At high $p_{T}$, the azimuthal angular correlation
changes little after charm quark pairs pass through the medium,
this is because the relaxation rate at high $p_{T}$ is small, see
Fig. \ref{fig:RelaxationRate}, so the angular deflection of the charm
quark is small in each scattering by thermal partons. At low $p_{T}$,
though a large angular deflection may occur in each scattering which
would reduce the angular correlation, since the original azimuthal angular correlation
is already very weak, such a reduction can hardly be observed. The
reason for such results is that a large default value $\left\langle k_{T}\right\rangle =2~\mathrm{GeV}$
for the Gaussian $k_{T}$ width of primordial partons in colliding
protons is used in PYTHIA 6.4 which dismisses the angular correlation
of low $p_{T}$ charm quark pairs. This is different from the result
of Ref. \cite{Zhu2008} which used momentum-independent transport
coefficients and a lower $\left\langle k_{T}\right\rangle $ value.
It is shown that if the thermal medium has large collective flow,
such as the partonic medium created at the LHC energy, the near side
instead of the back-to-back correlation of charm quark pairs would
appear. At such a high collisional energy, the next-to-leading order
PQCD processes for charm quark pairs become more important, and regeneration
of charm quarks in the partonic medium should also be taken into account
\cite{Uphoff2010}, which will further modify the angular correlation
of charm quark pairs at low $p_{T}$.

\begin{figure}
\begin{centering}
\includegraphics[scale=0.45]{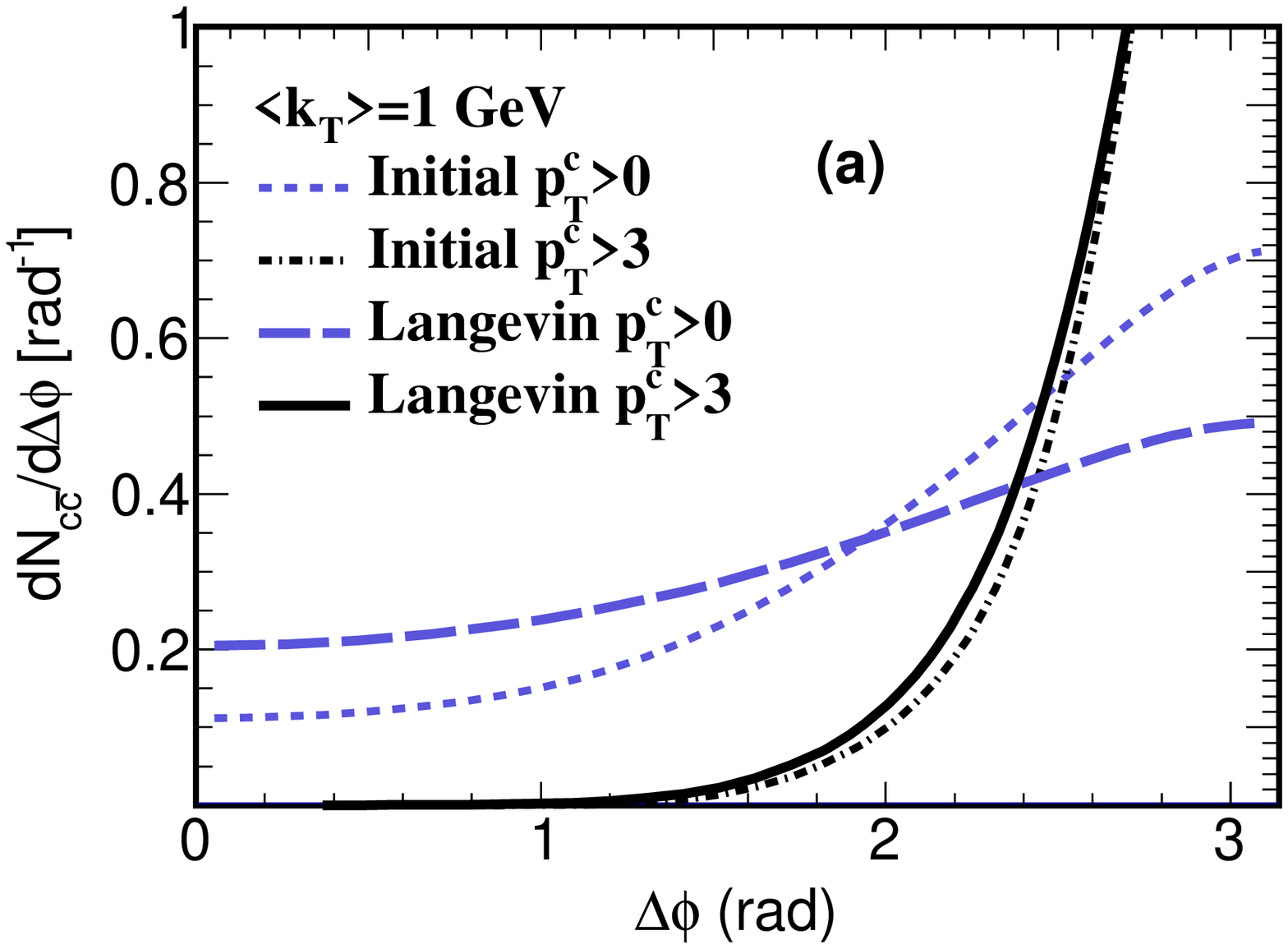}
\par\end{centering}

\begin{centering}
\includegraphics[scale=0.45]{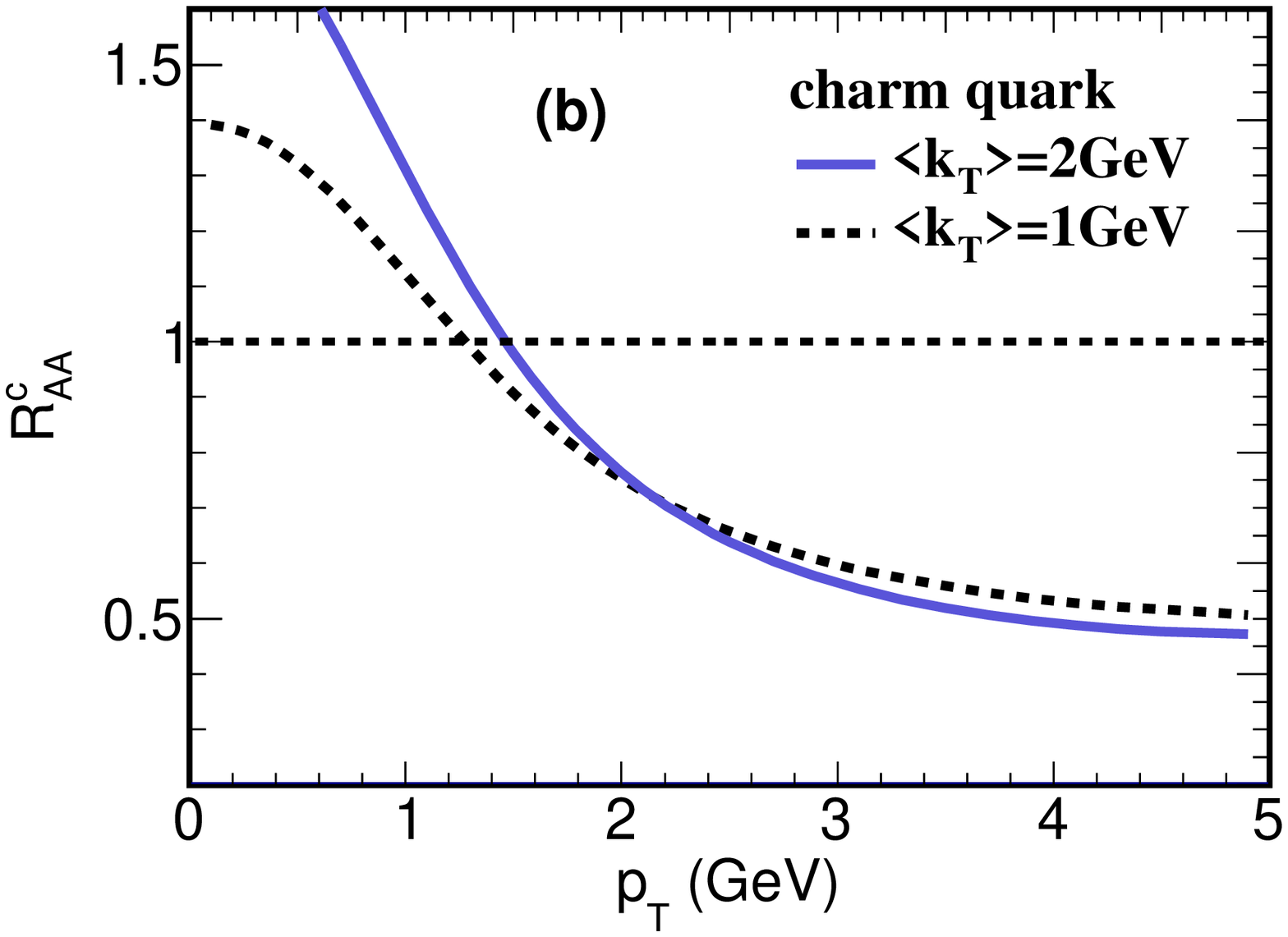}
\par\end{centering}

\caption{(Color online) (a) Same as Fig.~\ref{fig:charmquark}(b) but with $\left\langle k_{T}\right\rangle =1~\mathrm{GeV/c}$. The integrated multiplicities are normalized to unit.
(b) The nuclear modification factors with Hydro-Langevin evolution
for charm quarks in partonic medium for two values of $\left\langle k_{T}\right\rangle $
in PYTHIA. \label{fig:comparekT}}
\end{figure}

\subsubsection{Parameters dependance}
To look at the dependence of the correlation in azimuthal angles on $\left\langle k_{T}\right\rangle $,
we set $\left\langle k_{T}\right\rangle =1~\mathrm{GeV}$,
i.e. the default value of PYTHIA 6.3. This corresponds to a softer
$p_{T}$ distribution of charm quarks in the initial state. Then the
change of the azimuthal angular correlation at low $p_{T}$ is more obvious,
but the correlation at high $p_{T}$ is insensitive to the value of
$\left\langle k_{T}\right\rangle $. The nuclear modification factors
for two values of $\left\langle k_{T}\right\rangle $ in PYTHIA are
shown in Fig. \ref{fig:comparekT}(b).

\begin{figure}
\begin{centering}
\includegraphics[scale=0.45]{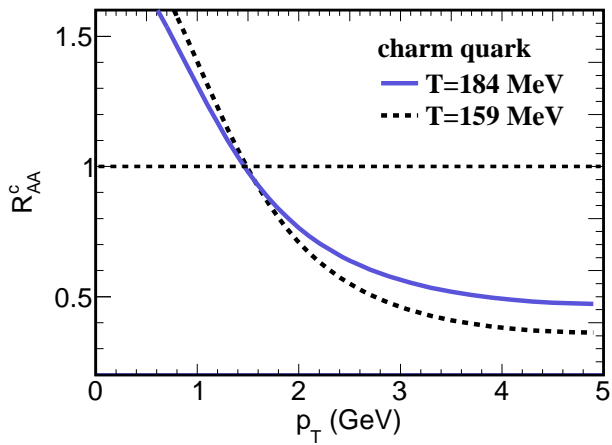}
\par\end{centering}

\caption{(Color online) Comparison the energy loss of charm quarks with different decoupling
temperature. The medium effect of charm quarks is given by the Hydro-Langevin
evolution. \label{fig:ComparsionT}}

\end{figure}

The freeze-out temperature $T_{c}$ influences the lifetime of the
partonic medium and then the degree of equilibrium of charm quarks.
To look at the $T_{c}$ dependence, we tune it to a lower value, $T_{c}=159$
MeV, which corresponds to energy density $e_{c}=0.445~\mathrm{GeV/fm^{3}}$.
The lower freeze-out temperature gives a longer evolution time of charm
quarks in the partonic medium and more interaction with thermal partons,
which give additional contribution to their energy loss, as shown
in Fig. \ref{fig:ComparsionT}. The diffusion effect of open charm
hadrons in hadronic medium can partially be accounted by lower freeze-out
temperature, in this sense our result is similar to a Langevin simulation
of open charm hadrons in hadronic medium \cite{Das:2013lra}.

\section{Experimental observables}
\label{sec:observable}
\subsection{Non-photonic electrons and di-electron background}
\begin{figure}
\begin{centering}
\includegraphics[scale=0.45]{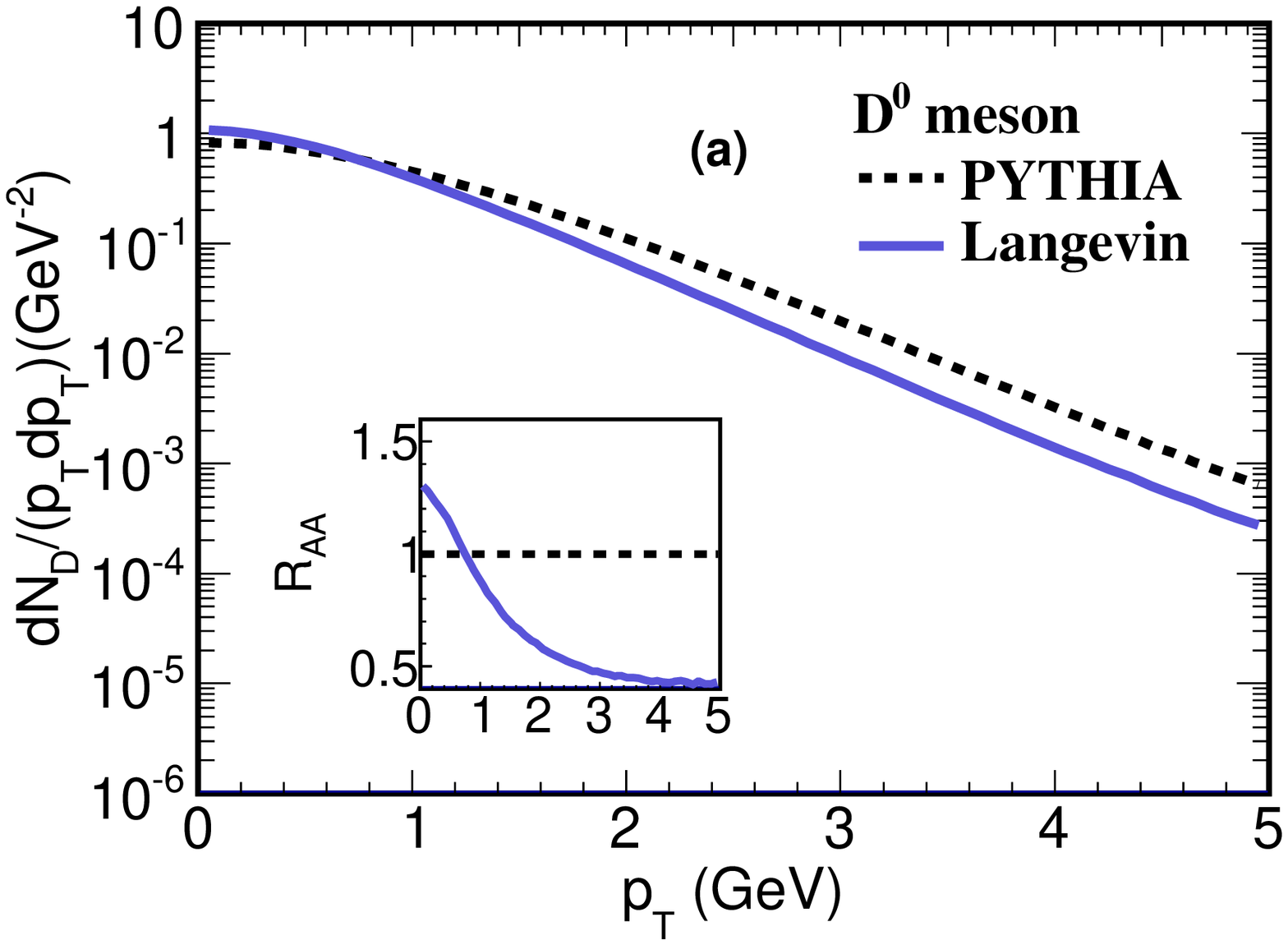}
\par\end{centering}

\begin{centering}
\includegraphics[scale=0.45]{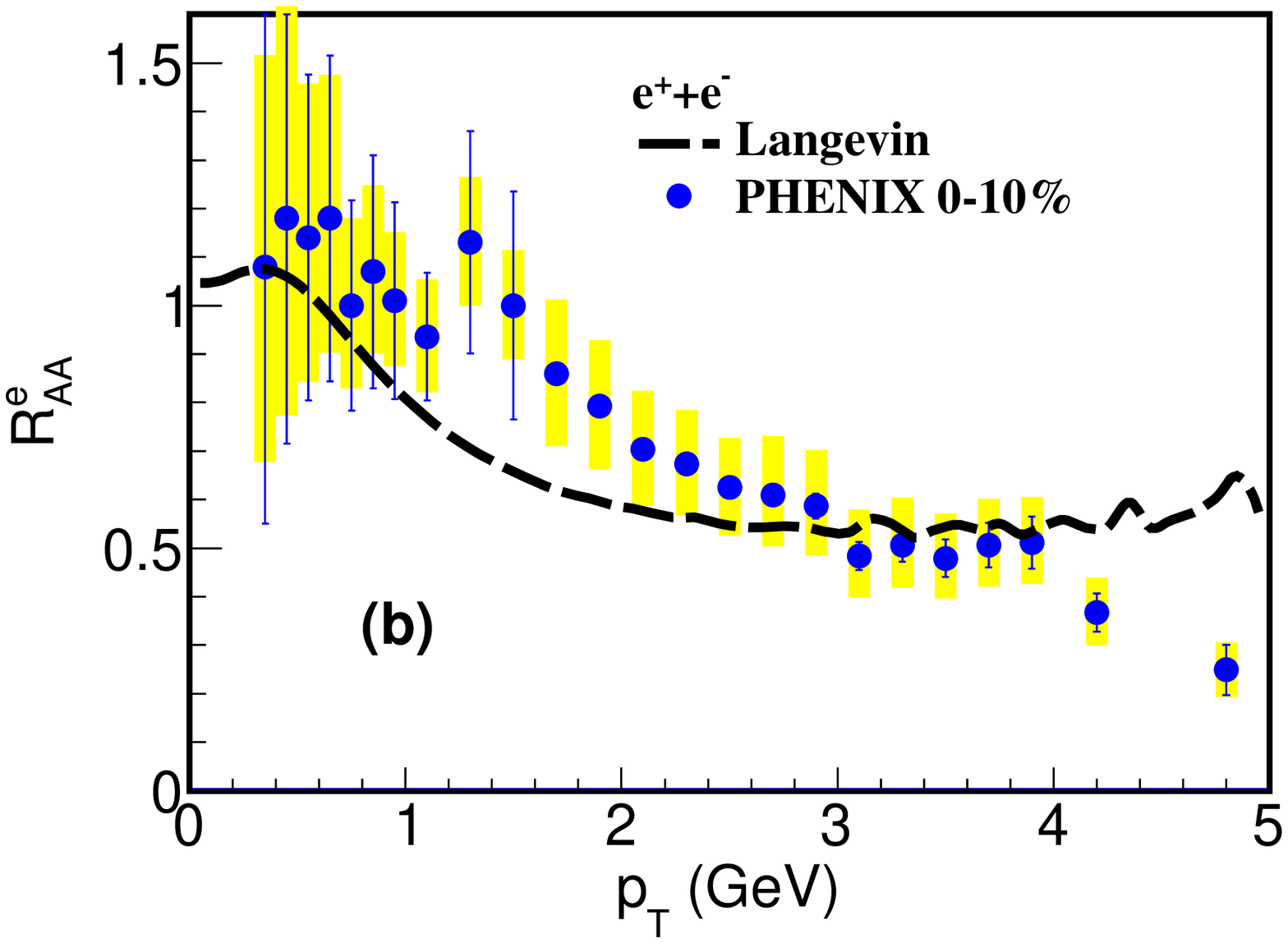}
\par\end{centering}

\caption{(Color online) (a) The $p_{T}$ spectra and the nuclear modification factor for $D^{0}$
mesons. The integrated multiplicity is normalized to unity. The nuclear modification factor is shown in the inset.
(b) The nuclear modification factor of electrons from semi-leptonic decays of open charm hadrons.
The data are taken from PHENIX \cite{Adare2011}.
\label{fig:semileptonicE}}

\end{figure}

After Hydro-Langevin evolution in the partonic
medium, charm quarks are put back to the same event in PYTHIA to undergo
hadronization and resonance decays. In Fig. \ref{fig:semileptonicE}(a),
we show the $p_{T}$ spectra and the nuclear modification factor of
$D^{0}$. The result of $R_{AA}$ for single electrons from
semi-leptonic decays of open charm hadrons is shown in Fig. \ref{fig:semileptonicE}(b),
in comparison with the PHENIX data in central Au+Au collisions \cite{Adare2011}.
Our result agrees with the data at intermediate $p_{T}$. In the low and high $p_{T}$ range,
the disagreement is large. In order to improve our result in this region,
we have to use the coalescence model to describe the hadronization of
the charm mesons and to include the contribution from bottom hadron decays.

\begin{figure}
\includegraphics[scale=0.45]{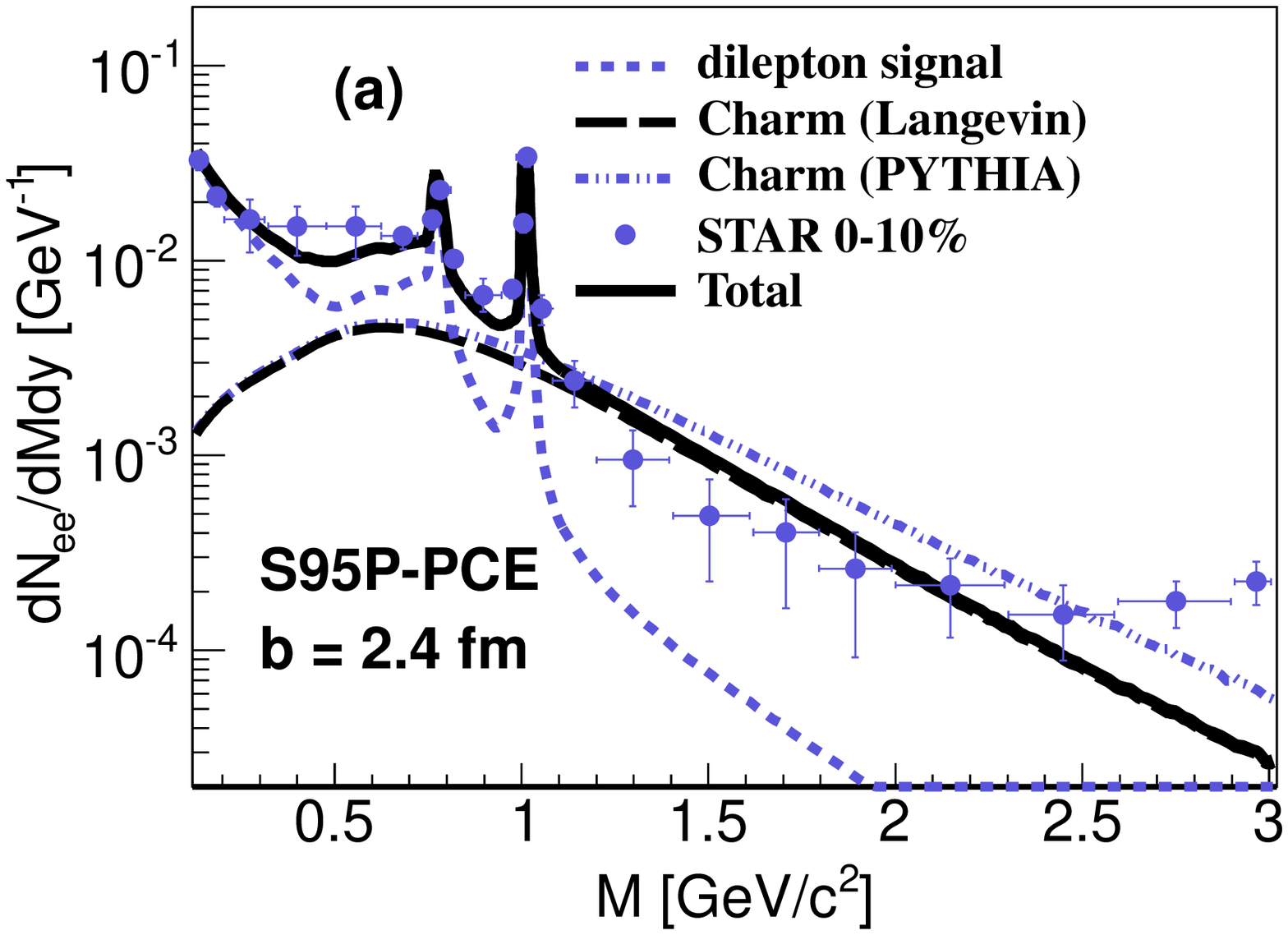}
\includegraphics[scale=0.45]{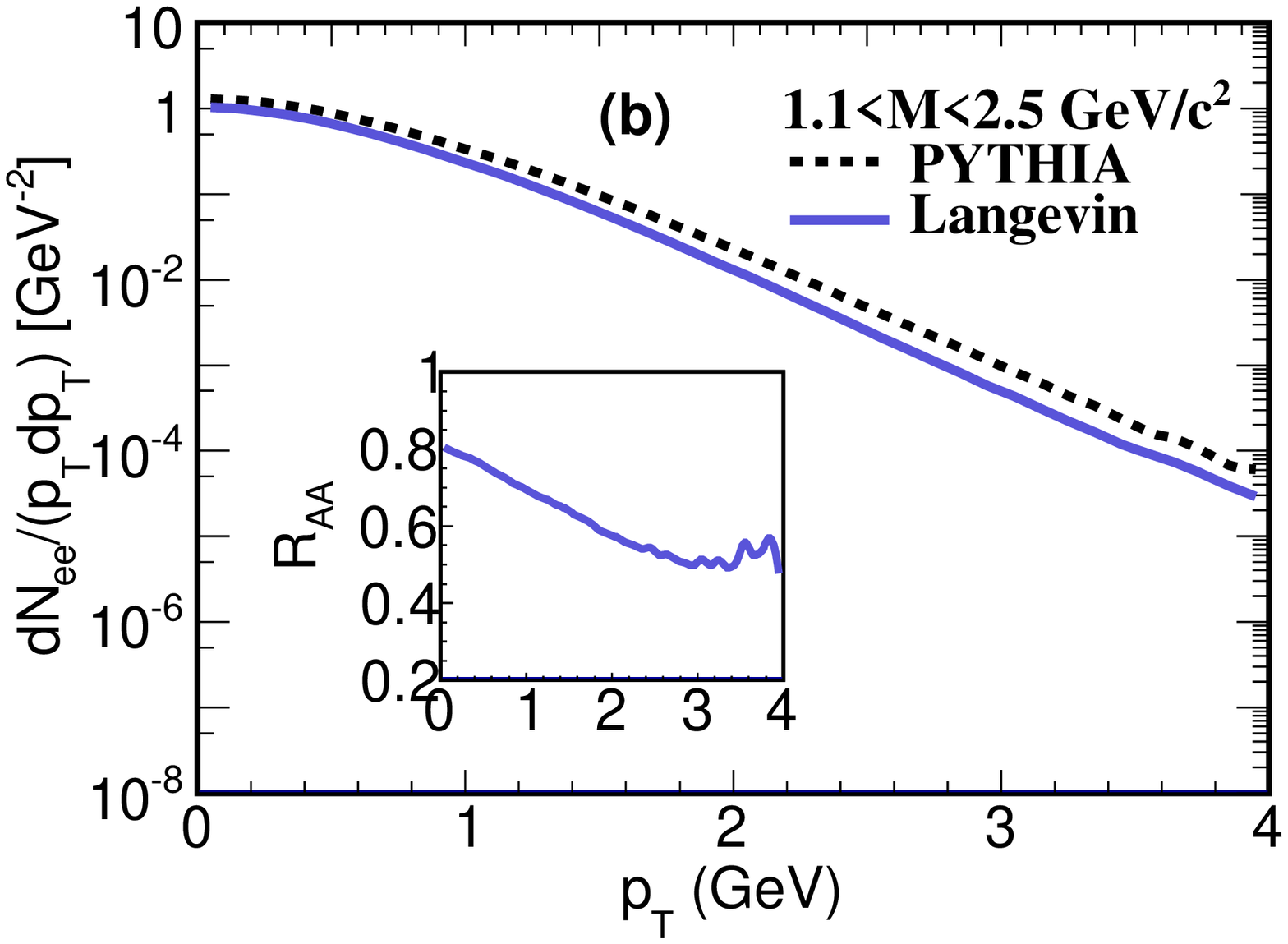}
\includegraphics[scale=0.45]{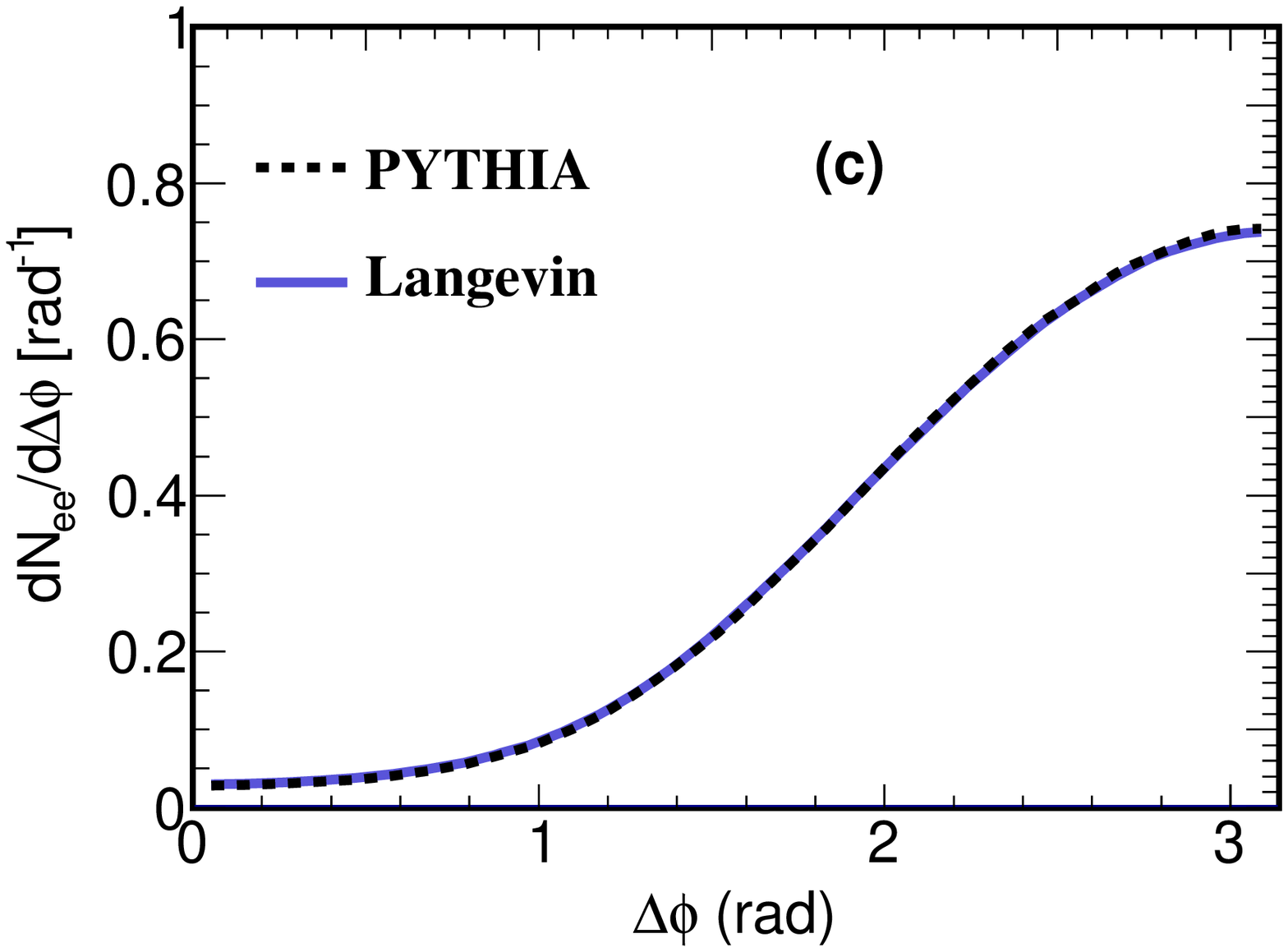}
\caption{(Color online) (a) Same as Fig.\ref{fig:ppcollision}(b) but the charm background is simulated by Hydro-Langevin
evolution. (b) The $p_{T}$ spectra and (c) The correlation in azimuthal angles for di-electrons from
semi-leptonic decays of correlated charm hadrons in the mass range $1.1<M<2.5~\mathrm{GeV/c^2}$. The nuclear modification factor of di-electrons is shown in the inset of (b).
The black-dashed line in (b) and (c) are the spectra from direct PYTHIA simulation with the same total yield. \label{fig:STARdilepton}}
\end{figure}

Now we come back to the di-electron invariant mass spectra as shown in Fig.~\ref{fig:ppcollision}(b) but using the charm background with medium modification from hydro-Langevin simulation. We see in Fig.~\ref{fig:STARdilepton}(a) that the medium effect suppresses the charm background in IMR and HMR compared to the PYTHIA result, and the medium modified charm background (black-long-dashed line) dominates over the signal (blue-dashed line) in the IMR. A better agreement with the data has been achieved. The contribution from open charm hadrons by hydro-Langevin simulation is larger (or less suppressed) in the IMR than our previous result \cite{Xu2012}.

In general, the suppression of the charm background in the IMR may be caused by
the charm quark energy loss or dismissing of the charm quark angular correlation.
In the mass range $1.1<M<2.5~\mathrm{GeV/c^2}$,
the $p_{T}$ spectra and the correlation in azimuthal angles for di-electrons
with the STAR detector acceptance are shown in Fig. \ref{fig:STARdilepton}(b,c).
The black-dashed line includes the open charm contribution without
medium modification whose spectra are the same as in p+p collisions
up to the normalization factor of the binary collision number. The blue-solid
line is the open charm contribution with medium modification
from the Hydro-Langevin evolution. One can see the suppression of
the $p_{T}$ spectra from the medium effect, however the correlation in azimuthal angles
is almost intact. This implies that the suppression of charm background in the IMR
is dominated by the charm quark energy loss. On the other hand, our current results
indicate that our previous naive model \cite{Xu2012} for the medium
modification of charm quarks, albeit simple, works well in the IMR.

\subsection{Remarks on charm quark hadronization}
In this paper, we use the Lund string fragmentation model encoded in PYTHIA
for the hadronization of charm quarks in p+p and Au+Au
collisions. However the hadronization mechanism in p+p and Au+Au
collisions can be very different in some $p_{T}$ range. It is known that
in the low $p_{T}$ range, the coalescence rather than fragmentation
is more important in heavy ion collisions \cite{Fries2003,Greco2003a,Greco2003}.
To make a schematic study in the coalescence scenario,
we use the space-time-integrated distribution function instead of the space-time-dependent distribution for charm and
light quarks. The light quark distribution at the freeze-out is given by the Cooper-Frye-type formula,
\begin{equation}
\frac{dN_q}{d^2p_Tdy}=\frac{g_i}{(2\pi)^3}\int_{\mathrm{T_f=184~\mathrm{MeV}}}d\Sigma_\mu p^\mu n_F(p\cdot u),
\end{equation}
where $\Sigma_\mu$ denotes the normal vector of the freeze-out hypersurface, $n_F$ is the Fermi distribution,
and $g_i$ is the degeneracy factor.

The $p_{T}$ distribution of a charm meson formed in the coalescence of a charm and a light quark
is given by \cite{Oh:2009zj}
\begin{eqnarray}
\frac{dN}{d^{2}\mathbf{p_{T}}}&=&g_{M}\frac{(2\sqrt{\pi}\sigma)^3}{V}\int d^2\mathbf{p_{1T}}d^2\mathbf{p_{2T}}\frac{dN_q}{d^2\mathbf{p^{q}_{T}}}\frac{dN_Q}{d^2\mathbf{p^{Q}_{T}}}\nonumber\\
&&\times \exp(-\mathbf{k}^2\sigma^2)\delta(\mathbf{p_T}-\mathbf{p^{q}_{T}}-\mathbf{p^{Q}_{T}}),
\end{eqnarray}
where $g_M$ is the statistical factor and $\mathbf{k}=(m_q\mathbf{\tilde{p}^Q}-m_Q\mathbf{\tilde{p}^q})/ (m_q+m_Q)$ with $m_{q,Q}$ being the quark masses and and $\mathbf{\tilde{p}^{q,Q}}$ being transverse momenta defined
in the center-of-mass frame of the charm meson.
The width parameter is given by $\sigma=1/\sqrt{\mu\omega}$ with $\mu=m_q m_Q/(m_q+m_Q)$ and $\omega=0.33~\mathrm{GeV}$~\cite{Oh:2009zj}.

\begin{figure}
\begin{centering}
\includegraphics[scale=0.42]{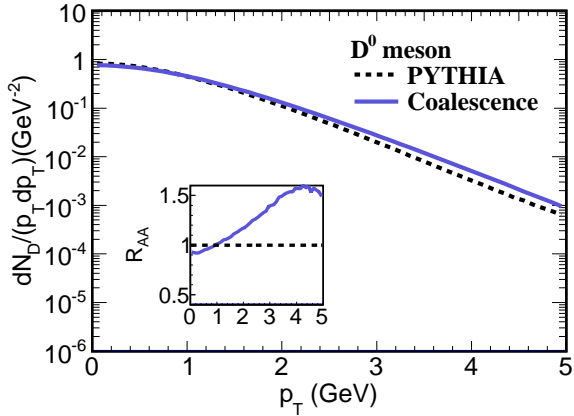}
\end{centering}
\caption{(Color online) The $p_{T}$ and $R_{AA}$ spectra for $D^{0}$
mesons from the coalescence model. The multiplicity is normalized to unity.
$R_{AA}$ is shown in the inset.
\label{fig:coalescence}}

\end{figure}

In Fig.~\ref{fig:coalescence} we show the $p_T$ and $R_{AA}$ spectra for $D_0$ mesons in coalescence hadronization scenario. We find that the $R_{AA}$ from the coalescence model is much larger than that from the fragmentation model. Since the coalescence mechanism dominates in the low $p_T$ region, we expect that the coalescence model could give a better description of single electron $R_{AA}$ at low $p_T$ than
Fig.~\ref{fig:semileptonicE}(b).
We will investigate the charm quark hadronization in details
in a separate paper.

In our work, we focus on the IMR di-electrons, most of which are from open charm hadron
decays with intermediate and high transverse momenta. This can be
seen by the obvious back-to-back azimuthal angular correlations of di-electrons
in Fig. \ref{fig:STARdilepton}(c). Given the uncertainties of coalescence
probability in the IMR, we expect that the modification from the coalescence/recombination
to the IMR di-electrons would be small.

\section{Summary and discussions}

\label{sec:Summary}
We investigate the medium modification of charm
quarks when they traverse the partonic medium using the relativistic
Fokker-Planck-Langevin equation for elastic scatterings of charm quarks
by thermal partons in a hydrodynamically expanding fireball. The transport
coefficients of charm quarks are calculated by the in-medium T-matrix
approximation with the static heavy quark potential, where the free
parameters are fitted by the color-average free energy of charm quarks
from LQCD. The space-time history of the fireball is provided by the
(2+1)-dimension ideal hydrodynamical model. We find that the suppression
of charm quark transverse momentum spectra depends on the transport
coefficients in the Langevin equation, the life-time of the partonic
medium, and the transverse momentum spectra of charm quarks
in the initial state. The medium modification of the correlation in azimuthal angles
of charm quark pairs turns out to be small in both the low and high
$p_{T}$ range, but it becomes larger in the low $p_{T}$ range when
a smaller $\left\langle k_{T}\right\rangle $ of primordial partons
in colliding protons is used in PYTHIA.

With hadronization of charm quarks, we calculate the $p_{T}$ spectra
and nuclear modification factor of $D^{0}$ mesons as well as the
nuclear modification factor of single electrons from open charm hadron
decays. Our results are in good agreement with the PHENIX data in
the intermediate $p_{T}$ range. With the STAR detector acceptance,
we compute the di-electron invariant mass spectra in most central
collisions and compare them with the STAR data of 0-10\%
centrality. Our current result is consistent to the previous one by
some of us \cite{Xu2012}. We find that the correlation in azimuthal angles for di-electrons
is almost identical to p+p and Au+Au collisions in the mass range
$1.1<M<2.5~\mathrm{GeV/c^2}$. This implies the suppression of
the charm background in the IMR is dominated by the charm quark
energy loss.

We try to give a quantitative study of the open charm contribution
to di-electrons in the IMR.
There are still some uncertainties which could be constrained
in a future more comprehensive study of open charm hadrons
in heavy ion collisions.
Firstly the initial charm quark distribution in p+p and Au+Au collisions
has to be fixed. There is a recent measurement of
the charm meson production cross section and
transverse momentum spectra \cite{Adamczyk2012} which may shed light
on this, but the statistics is not high enough and needs to be improved.
Secondly the open charm hadron diffusion in the hadronic medium might be
relevant. Though this effect can be partially achieved by tuning the
freeze-out temperature to lower values, a rigorous study is necessary.
Thirdly the charm quark hadronization has to be treated in a better
model. Though we argue that the hadronization process may not modify
our current IMR di-electron results very much, it is expected to have impact on
the low and intermediate $p_{T}$ region. Finally the charm quark
regeneration in the partonic medium may also be important at the LHC
energy.

\begin{acknowledgments}
HX thanks Y.-K. Song, B.-C. Huang and J. Song for helpful discussions.
QW is supported by National Natural Science Foundation of China under
the grant no. 11125524. This work was supported in part by the Offices
of NP and HEP within the U.S. DOE Office of Science under the contracts
of DE-FG02-88ER40412 and DE-AC02-98CH10886.USTC.
\end{acknowledgments}

\bibliographystyle{apsrev}
\addcontentsline{toc}{section}{\refname}\bibliography{HQref}

\begin{thebibliography}{45}
\expandafter\ifx\csname natexlab\endcsname\relax\def\natexlab#1{#1}\fi
\expandafter\ifx\csname bibnamefont\endcsname\relax
  \def\bibnamefont#1{#1}\fi
\expandafter\ifx\csname bibfnamefont\endcsname\relax
  \def\bibfnamefont#1{#1}\fi
\expandafter\ifx\csname citenamefont\endcsname\relax
  \def\citenamefont#1{#1}\fi
\expandafter\ifx\csname url\endcsname\relax
  \def\url#1{\texttt{#1}}\fi
\expandafter\ifx\csname urlprefix\endcsname\relax\def\urlprefix{URL }\fi
\providecommand{\bibinfo}[2]{#2}
\providecommand{\eprint}[2][]{\url{#2}}

\bibitem[{\citenamefont{Adams et~al.}(2005)}]{Adams:2005dq}
\bibinfo{author}{\bibfnamefont{J.}~\bibnamefont{Adams}} \bibnamefont{et~al.}
  (\bibinfo{collaboration}{STAR Collaboration}), \bibinfo{journal}{Nucl.Phys.}
  \textbf{\bibinfo{volume}{A757}}, \bibinfo{pages}{102} (\bibinfo{year}{2005}).

\bibitem[{\citenamefont{Adcox et~al.}(2005)}]{Adcox:2004mh}
\bibinfo{author}{\bibfnamefont{K.}~\bibnamefont{Adcox}} \bibnamefont{et~al.}
  (\bibinfo{collaboration}{PHENIX Collaboration}),
  \bibinfo{journal}{Nucl.Phys.} \textbf{\bibinfo{volume}{A757}},
  \bibinfo{pages}{184} (\bibinfo{year}{2005}).

\bibitem[{\citenamefont{Evans and Bryant}(2008)}]{Evans:2008zzb}
\bibinfo{author}{\bibfnamefont{L.}~\bibnamefont{Evans}} \bibnamefont{and}
  \bibinfo{author}{\bibfnamefont{P.}~\bibnamefont{Bryant}},
  \bibinfo{journal}{JINST} \textbf{\bibinfo{volume}{3}},
  \bibinfo{pages}{S08001} (\bibinfo{year}{2008}).

\bibitem[{\citenamefont{Ackermann et~al.}(2001)}]{Ackermann2001}
\bibinfo{author}{\bibfnamefont{K.}~\bibnamefont{Ackermann}}
  \bibnamefont{et~al.} (\bibinfo{collaboration}{STAR Collaboration}),
  \bibinfo{journal}{Phys.Rev.Lett.} \textbf{\bibinfo{volume}{86}},
  \bibinfo{pages}{402} (\bibinfo{year}{2001}).

\bibitem[{\citenamefont{Adcox et~al.}(2002)}]{Adcox2002}
\bibinfo{author}{\bibfnamefont{K.}~\bibnamefont{Adcox}} \bibnamefont{et~al.}
  (\bibinfo{collaboration}{PHENIX Collaboration}),
  \bibinfo{journal}{Phys.Rev.Lett.} \textbf{\bibinfo{volume}{89}},
  \bibinfo{pages}{212301} (\bibinfo{year}{2002}).

\bibitem[{\citenamefont{Aamodt et~al.}(2010)}]{Aamodt:2010pa}
\bibinfo{author}{\bibfnamefont{K.}~\bibnamefont{Aamodt}} \bibnamefont{et~al.}
  (\bibinfo{collaboration}{ALICE Collaboration}),
  \bibinfo{journal}{Phys.Rev.Lett.} \textbf{\bibinfo{volume}{105}},
  \bibinfo{pages}{252302} (\bibinfo{year}{2010}).

\bibitem[{\citenamefont{Gyulassy and McLerran}(2005)}]{Gyulassy:2004zy}
\bibinfo{author}{\bibfnamefont{M.}~\bibnamefont{Gyulassy}} \bibnamefont{and}
  \bibinfo{author}{\bibfnamefont{L.}~\bibnamefont{McLerran}},
  \bibinfo{journal}{Nucl.Phys.} \textbf{\bibinfo{volume}{A750}},
  \bibinfo{pages}{30} (\bibinfo{year}{2005}).

\bibitem[{\citenamefont{Shuryak}(2009)}]{Shuryak:2008eq}
\bibinfo{author}{\bibfnamefont{E.}~\bibnamefont{Shuryak}},
  \bibinfo{journal}{Prog.Part.Nucl.Phys.} \textbf{\bibinfo{volume}{62}},
  \bibinfo{pages}{48} (\bibinfo{year}{2009}).

\bibitem[{\citenamefont{McLerran and Toimela}(1985)}]{McLerran1985a}
\bibinfo{author}{\bibfnamefont{L.~D.} \bibnamefont{McLerran}} \bibnamefont{and}
  \bibinfo{author}{\bibfnamefont{T.}~\bibnamefont{Toimela}},
  \bibinfo{journal}{Phys. Rev.} \textbf{\bibinfo{volume}{D31}},
  \bibinfo{pages}{545} (\bibinfo{year}{1985}).

\bibitem[{\citenamefont{Kajantie et~al.}(1986)\citenamefont{Kajantie, Kapusta,
  McLerran, and Mekjian}}]{Kajantie1986dh}
\bibinfo{author}{\bibfnamefont{K.}~\bibnamefont{Kajantie}},
  \bibinfo{author}{\bibfnamefont{J.~I.} \bibnamefont{Kapusta}},
  \bibinfo{author}{\bibfnamefont{L.~D.} \bibnamefont{McLerran}},
  \bibnamefont{and} \bibinfo{author}{\bibfnamefont{A.}~\bibnamefont{Mekjian}},
  \bibinfo{journal}{Phys. Rev.} \textbf{\bibinfo{volume}{D34}},
  \bibinfo{pages}{2746} (\bibinfo{year}{1986}).

\bibitem[{\citenamefont{van Hees and Rapp}(2006)}]{vanHees:2006ng}
\bibinfo{author}{\bibfnamefont{H.}~\bibnamefont{van Hees}} \bibnamefont{and}
  \bibinfo{author}{\bibfnamefont{R.}~\bibnamefont{Rapp}},
  \bibinfo{journal}{Phys. Rev. Lett.} \textbf{\bibinfo{volume}{97}},
  \bibinfo{pages}{102301} (\bibinfo{year}{2006}).

\bibitem[{\citenamefont{Ruppert et~al.}(2008)\citenamefont{Ruppert, Gale, Renk,
  Lichard, and Kapusta}}]{Ruppert2008}
\bibinfo{author}{\bibfnamefont{J.}~\bibnamefont{Ruppert}},
  \bibinfo{author}{\bibfnamefont{C.}~\bibnamefont{Gale}},
  \bibinfo{author}{\bibfnamefont{T.}~\bibnamefont{Renk}},
  \bibinfo{author}{\bibfnamefont{P.}~\bibnamefont{Lichard}}, \bibnamefont{and}
  \bibinfo{author}{\bibfnamefont{J.~I.} \bibnamefont{Kapusta}},
  \bibinfo{journal}{Phys.Rev.Lett.} \textbf{\bibinfo{volume}{100}},
  \bibinfo{pages}{162301} (\bibinfo{year}{2008}).

\bibitem[{\citenamefont{Dusling et~al.}(2007)\citenamefont{Dusling, Teaney, and
  Zahed}}]{Dusling2007}
\bibinfo{author}{\bibfnamefont{K.}~\bibnamefont{Dusling}},
  \bibinfo{author}{\bibfnamefont{D.}~\bibnamefont{Teaney}}, \bibnamefont{and}
  \bibinfo{author}{\bibfnamefont{I.}~\bibnamefont{Zahed}},
  \bibinfo{journal}{Phys.Rev.} \textbf{\bibinfo{volume}{C75}},
  \bibinfo{pages}{024908} (\bibinfo{year}{2007}).

\bibitem[{\citenamefont{Xu et~al.}(2012)\citenamefont{Xu, Chen, Dong, Wang, and
  Zhang}}]{Xu2012}
\bibinfo{author}{\bibfnamefont{H.-j.} \bibnamefont{Xu}},
  \bibinfo{author}{\bibfnamefont{H.-f.} \bibnamefont{Chen}},
  \bibinfo{author}{\bibfnamefont{X.}~\bibnamefont{Dong}},
  \bibinfo{author}{\bibfnamefont{Q.}~\bibnamefont{Wang}}, \bibnamefont{and}
  \bibinfo{author}{\bibfnamefont{Y.-f.} \bibnamefont{Zhang}},
  \bibinfo{journal}{Phys.Rev.} \textbf{\bibinfo{volume}{C85}},
  \bibinfo{pages}{024906} (\bibinfo{year}{2012}).

\bibitem[{\citenamefont{Linnyk et~al.}(2012)\citenamefont{Linnyk, Cassing,
  Manninen, Bratkovskaya, and Ko}}]{Linnyk:2011vx}
\bibinfo{author}{\bibfnamefont{O.}~\bibnamefont{Linnyk}},
  \bibinfo{author}{\bibfnamefont{W.}~\bibnamefont{Cassing}},
  \bibinfo{author}{\bibfnamefont{J.}~\bibnamefont{Manninen}},
  \bibinfo{author}{\bibfnamefont{E.}~\bibnamefont{Bratkovskaya}},
  \bibnamefont{and} \bibinfo{author}{\bibfnamefont{C.}~\bibnamefont{Ko}},
  \bibinfo{journal}{Phys.Rev.} \textbf{\bibinfo{volume}{C85}},
  \bibinfo{pages}{024910} (\bibinfo{year}{2012}).

\bibitem[{\citenamefont{Deng et~al.}(2011)\citenamefont{Deng, Wang, Xu, and
  Zhuang}}]{Deng2011}
\bibinfo{author}{\bibfnamefont{J.}~\bibnamefont{Deng}},
  \bibinfo{author}{\bibfnamefont{Q.}~\bibnamefont{Wang}},
  \bibinfo{author}{\bibfnamefont{N.}~\bibnamefont{Xu}}, \bibnamefont{and}
  \bibinfo{author}{\bibfnamefont{P.}~\bibnamefont{Zhuang}},
  \bibinfo{journal}{Phys.Lett.} \textbf{\bibinfo{volume}{B701}},
  \bibinfo{pages}{581} (\bibinfo{year}{2011}).

\bibitem[{\citenamefont{Wicks et~al.}(2007)\citenamefont{Wicks, Horowitz,
  Djordjevic, and Gyulassy}}]{Wicks2007}
\bibinfo{author}{\bibfnamefont{S.}~\bibnamefont{Wicks}},
  \bibinfo{author}{\bibfnamefont{W.}~\bibnamefont{Horowitz}},
  \bibinfo{author}{\bibfnamefont{M.}~\bibnamefont{Djordjevic}},
  \bibnamefont{and} \bibinfo{author}{\bibfnamefont{M.}~\bibnamefont{Gyulassy}},
  \bibinfo{journal}{Nucl.Phys.} \textbf{\bibinfo{volume}{A784}},
  \bibinfo{pages}{426} (\bibinfo{year}{2007}).

\bibitem[{\citenamefont{Dokshitzer and Kharzeev}(2001)}]{Dokshitzer2001}
\bibinfo{author}{\bibfnamefont{Y.~L.} \bibnamefont{Dokshitzer}}
  \bibnamefont{and} \bibinfo{author}{\bibfnamefont{D.}~\bibnamefont{Kharzeev}},
  \bibinfo{journal}{Phys.Lett.} \textbf{\bibinfo{volume}{B519}},
  \bibinfo{pages}{199} (\bibinfo{year}{2001}).

\bibitem[{\citenamefont{Adare et~al.}(2007)}]{Adare2007}
\bibinfo{author}{\bibfnamefont{A.}~\bibnamefont{Adare}} \bibnamefont{et~al.}
  (\bibinfo{collaboration}{PHENIX Collaboration}),
  \bibinfo{journal}{Phys.Rev.Lett.} \textbf{\bibinfo{volume}{98}},
  \bibinfo{pages}{172301} (\bibinfo{year}{2007}).

\bibitem[{\citenamefont{Alberico et~al.}(2011)\citenamefont{Alberico, Beraudo,
  De~Pace, Molinari, Monteno et~al.}}]{Alberico2011c}
\bibinfo{author}{\bibfnamefont{W.}~\bibnamefont{Alberico}},
  \bibinfo{author}{\bibfnamefont{A.}~\bibnamefont{Beraudo}},
  \bibinfo{author}{\bibfnamefont{A.}~\bibnamefont{De~Pace}},
  \bibinfo{author}{\bibfnamefont{A.}~\bibnamefont{Molinari}},
  \bibinfo{author}{\bibfnamefont{M.}~\bibnamefont{Monteno}},
  \bibnamefont{et~al.}, \bibinfo{journal}{Eur.Phys.J.}
  \textbf{\bibinfo{volume}{C71}}, \bibinfo{pages}{1666} (\bibinfo{year}{2011}).

\bibitem[{\citenamefont{van Hees and Rapp}(2005)}]{Hees2005}
\bibinfo{author}{\bibfnamefont{H.}~\bibnamefont{van Hees}} \bibnamefont{and}
  \bibinfo{author}{\bibfnamefont{R.}~\bibnamefont{Rapp}},
  \bibinfo{journal}{Phys.Rev.} \textbf{\bibinfo{volume}{C71}},
  \bibinfo{pages}{034907} (\bibinfo{year}{2005}).

\bibitem[{\citenamefont{van Hees et~al.}(2008)\citenamefont{van Hees,
  Mannarelli, Greco, and Rapp}}]{Hees2008}
\bibinfo{author}{\bibfnamefont{H.}~\bibnamefont{van Hees}},
  \bibinfo{author}{\bibfnamefont{M.}~\bibnamefont{Mannarelli}},
  \bibinfo{author}{\bibfnamefont{V.}~\bibnamefont{Greco}}, \bibnamefont{and}
  \bibinfo{author}{\bibfnamefont{R.}~\bibnamefont{Rapp}},
  \bibinfo{journal}{Phys.Rev.Lett.} \textbf{\bibinfo{volume}{100}},
  \bibinfo{pages}{192301} (\bibinfo{year}{2008}).

\bibitem[{\citenamefont{He et~al.}(2012)\citenamefont{He, Fries, and
  Rapp}}]{He2012}
\bibinfo{author}{\bibfnamefont{M.}~\bibnamefont{He}},
  \bibinfo{author}{\bibfnamefont{R.~J.} \bibnamefont{Fries}}, \bibnamefont{and}
  \bibinfo{author}{\bibfnamefont{R.}~\bibnamefont{Rapp}},
  \bibinfo{journal}{Phys.Rev.} \textbf{\bibinfo{volume}{C86}},
  \bibinfo{pages}{014903} (\bibinfo{year}{2012}).

\bibitem[{\citenamefont{Uphoff et~al.}(2010)\citenamefont{Uphoff, Fochler, Xu,
  and Greiner}}]{Uphoff2010}
\bibinfo{author}{\bibfnamefont{J.}~\bibnamefont{Uphoff}},
  \bibinfo{author}{\bibfnamefont{O.}~\bibnamefont{Fochler}},
  \bibinfo{author}{\bibfnamefont{Z.}~\bibnamefont{Xu}}, \bibnamefont{and}
  \bibinfo{author}{\bibfnamefont{C.}~\bibnamefont{Greiner}},
  \bibinfo{journal}{Phys.Rev.} \textbf{\bibinfo{volume}{C82}},
  \bibinfo{pages}{044906} (\bibinfo{year}{2010}).

\bibitem[{\citenamefont{Uphoff et~al.}(2012)\citenamefont{Uphoff, Fochler, Xu,
  and Greiner}}]{Uphoff2012}
\bibinfo{author}{\bibfnamefont{J.}~\bibnamefont{Uphoff}},
  \bibinfo{author}{\bibfnamefont{O.}~\bibnamefont{Fochler}},
  \bibinfo{author}{\bibfnamefont{Z.}~\bibnamefont{Xu}}, \bibnamefont{and}
  \bibinfo{author}{\bibfnamefont{C.}~\bibnamefont{Greiner}},
  \bibinfo{journal}{Phys.Lett.} \textbf{\bibinfo{volume}{B717}},
  \bibinfo{pages}{430} (\bibinfo{year}{2012}).

\bibitem[{\citenamefont{Mannarelli and Rapp}(2005)}]{Mannarelli2005}
\bibinfo{author}{\bibfnamefont{M.}~\bibnamefont{Mannarelli}} \bibnamefont{and}
  \bibinfo{author}{\bibfnamefont{R.}~\bibnamefont{Rapp}},
  \bibinfo{journal}{Phys.Rev.} \textbf{\bibinfo{volume}{C72}},
  \bibinfo{pages}{064905} (\bibinfo{year}{2005}).

\bibitem[{\citenamefont{Riek and Rapp}(2010)}]{Riek2010}
\bibinfo{author}{\bibfnamefont{F.}~\bibnamefont{Riek}} \bibnamefont{and}
  \bibinfo{author}{\bibfnamefont{R.}~\bibnamefont{Rapp}},
  \bibinfo{journal}{Phys.Rev.} \textbf{\bibinfo{volume}{C82}},
  \bibinfo{pages}{035201} (\bibinfo{year}{2010}).

\bibitem[{\citenamefont{Huggins and Rapp}(2012)}]{Huggins2012}
\bibinfo{author}{\bibfnamefont{K.}~\bibnamefont{Huggins}} \bibnamefont{and}
  \bibinfo{author}{\bibfnamefont{R.}~\bibnamefont{Rapp}},
  \bibinfo{journal}{Nucl.Phys.} \textbf{\bibinfo{volume}{A896}},
  \bibinfo{pages}{24} (\bibinfo{year}{2012}).

\bibitem[{\citenamefont{Zhu et~al.}(2007)\citenamefont{Zhu, Bleicher, Huang,
  Schweda, Stoecker et~al.}}]{Zhu2007}
\bibinfo{author}{\bibfnamefont{X.}~\bibnamefont{Zhu}},
  \bibinfo{author}{\bibfnamefont{M.}~\bibnamefont{Bleicher}},
  \bibinfo{author}{\bibfnamefont{S.}~\bibnamefont{Huang}},
  \bibinfo{author}{\bibfnamefont{K.}~\bibnamefont{Schweda}},
  \bibinfo{author}{\bibfnamefont{H.}~\bibnamefont{Stoecker}},
  \bibnamefont{et~al.}, \bibinfo{journal}{Phys.Lett.}
  \textbf{\bibinfo{volume}{B647}}, \bibinfo{pages}{366} (\bibinfo{year}{2007}).

\bibitem[{\citenamefont{Zhu et~al.}(2008)\citenamefont{Zhu, Xu, and
  Zhuang}}]{Zhu2008}
\bibinfo{author}{\bibfnamefont{X.}~\bibnamefont{Zhu}},
  \bibinfo{author}{\bibfnamefont{N.}~\bibnamefont{Xu}}, \bibnamefont{and}
  \bibinfo{author}{\bibfnamefont{P.}~\bibnamefont{Zhuang}},
  \bibinfo{journal}{Phys.Rev.Lett.} \textbf{\bibinfo{volume}{100}},
  \bibinfo{pages}{152301} (\bibinfo{year}{2008}).

\bibitem[{\citenamefont{Sjostrand et~al.}(2001)}]{Sjostrand:2000wi}
\bibinfo{author}{\bibfnamefont{T.}~\bibnamefont{Sjostrand}}
  \bibnamefont{et~al.}, \bibinfo{journal}{Comput. Phys. Commun.}
  \textbf{\bibinfo{volume}{135}}, \bibinfo{pages}{238} (\bibinfo{year}{2001}).

\bibitem[{\citenamefont{Adare et~al.}(2009)}]{Adare2009}
\bibinfo{author}{\bibfnamefont{A.}~\bibnamefont{Adare}} \bibnamefont{et~al.}
  (\bibinfo{collaboration}{PHENIX Collaboration}),
  \bibinfo{journal}{Phys.Lett.} \textbf{\bibinfo{volume}{B670}},
  \bibinfo{pages}{313} (\bibinfo{year}{2009}).

\bibitem[{\citenamefont{Eidelman et~al.}(2004)}]{Eidelman2004}
\bibinfo{author}{\bibfnamefont{S.}~\bibnamefont{Eidelman}} \bibnamefont{et~al.}
  (\bibinfo{collaboration}{Particle Data Group}), \bibinfo{journal}{Phys.Lett.}
  \textbf{\bibinfo{volume}{B592}}, \bibinfo{pages}{1} (\bibinfo{year}{2004}).

\bibitem[{\citenamefont{Zhao}(2011)}]{Zhao2011}
\bibinfo{author}{\bibfnamefont{J.}~\bibnamefont{Zhao}}
  (\bibinfo{collaboration}{STAR Collaboration}), \bibinfo{journal}{J.Phys.}
  \textbf{\bibinfo{volume}{G38}}, \bibinfo{pages}{124134}
  (\bibinfo{year}{2011}).

\bibitem[{\citenamefont{Adamczyk et~al.}(2013)}]{Adamczyk:2013caa}
\bibinfo{author}{\bibfnamefont{L.}~\bibnamefont{Adamczyk}} \bibnamefont{et~al.}
  (\bibinfo{collaboration}{STAR Collaboration}) (\bibinfo{year}{2013}),
  \eprint{1312.7397}.

\bibitem[{\citenamefont{Svetitsky}(1988)}]{Svetitsky1988}
\bibinfo{author}{\bibfnamefont{B.}~\bibnamefont{Svetitsky}},
  \bibinfo{journal}{Phys.Rev.} \textbf{\bibinfo{volume}{D37}},
  \bibinfo{pages}{2484} (\bibinfo{year}{1988}).

\bibitem[{\citenamefont{Dunkel and Hänggi}(2009)}]{Dunkel2009a}
\bibinfo{author}{\bibfnamefont{J.}~\bibnamefont{Dunkel}} \bibnamefont{and}
  \bibinfo{author}{\bibfnamefont{P.}~\bibnamefont{Hänggi}},
  \bibinfo{journal}{Phys.Rep.} \textbf{\bibinfo{volume}{471}},
  \bibinfo{pages}{1} (\bibinfo{year}{2009}).

\bibitem[{\citenamefont{Das et~al.}(2013)\citenamefont{Das, Ghosh, Sarkar, and
  Alam}}]{Das:2013lra}
\bibinfo{author}{\bibfnamefont{S.~K.} \bibnamefont{Das}},
  \bibinfo{author}{\bibfnamefont{S.}~\bibnamefont{Ghosh}},
  \bibinfo{author}{\bibfnamefont{S.}~\bibnamefont{Sarkar}}, \bibnamefont{and}
  \bibinfo{author}{\bibfnamefont{J.-e.} \bibnamefont{Alam}}
  (\bibinfo{year}{2013}).

\bibitem[{\citenamefont{Dunkel et~al.}(2009)\citenamefont{Dunkel, Hanggi, and
  Weber}}]{Dunkel2009}
\bibinfo{author}{\bibfnamefont{J.}~\bibnamefont{Dunkel}},
  \bibinfo{author}{\bibfnamefont{P.}~\bibnamefont{Hanggi}}, \bibnamefont{and}
  \bibinfo{author}{\bibfnamefont{S.}~\bibnamefont{Weber}},
  \bibinfo{journal}{Phys.Rev.} \textbf{\bibinfo{volume}{E79}},
  \bibinfo{pages}{010101} (\bibinfo{year}{2009}).

\bibitem[{\citenamefont{Adare et~al.}(2011)}]{Adare2011}
\bibinfo{author}{\bibfnamefont{A.}~\bibnamefont{Adare}} \bibnamefont{et~al.}
  (\bibinfo{collaboration}{PHENIX Collaboration}), \bibinfo{journal}{Phys.Rev.}
  \textbf{\bibinfo{volume}{C84}}, \bibinfo{pages}{044905}
  (\bibinfo{year}{2011}).

\bibitem[{\citenamefont{Fries et~al.}(2003)\citenamefont{Fries, Muller, Nonaka,
  and Bass}}]{Fries2003}
\bibinfo{author}{\bibfnamefont{R.}~\bibnamefont{Fries}},
  \bibinfo{author}{\bibfnamefont{B.}~\bibnamefont{Muller}},
  \bibinfo{author}{\bibfnamefont{C.}~\bibnamefont{Nonaka}}, \bibnamefont{and}
  \bibinfo{author}{\bibfnamefont{S.}~\bibnamefont{Bass}},
  \bibinfo{journal}{Phys.Rev.Lett.} \textbf{\bibinfo{volume}{90}},
  \bibinfo{pages}{202303} (\bibinfo{year}{2003}).

\bibitem[{\citenamefont{Greco et~al.}(2003{\natexlab{a}})\citenamefont{Greco,
  Ko, and Levai}}]{Greco2003a}
\bibinfo{author}{\bibfnamefont{V.}~\bibnamefont{Greco}},
  \bibinfo{author}{\bibfnamefont{C.}~\bibnamefont{Ko}}, \bibnamefont{and}
  \bibinfo{author}{\bibfnamefont{P.}~\bibnamefont{Levai}},
  \bibinfo{journal}{Phys.Rev.} \textbf{\bibinfo{volume}{C68}},
  \bibinfo{pages}{034904} (\bibinfo{year}{2003}{\natexlab{a}}).

\bibitem[{\citenamefont{Greco et~al.}(2003{\natexlab{b}})\citenamefont{Greco,
  Ko, and Levai}}]{Greco2003}
\bibinfo{author}{\bibfnamefont{V.}~\bibnamefont{Greco}},
  \bibinfo{author}{\bibfnamefont{C.}~\bibnamefont{Ko}}, \bibnamefont{and}
  \bibinfo{author}{\bibfnamefont{P.}~\bibnamefont{Levai}},
  \bibinfo{journal}{Phys.Rev.Lett.} \textbf{\bibinfo{volume}{90}},
  \bibinfo{pages}{202302} (\bibinfo{year}{2003}{\natexlab{b}}).

\bibitem[{\citenamefont{Oh et~al.}(2009)\citenamefont{Oh, Ko, Lee, and
  Yasui}}]{Oh:2009zj}
\bibinfo{author}{\bibfnamefont{Y.}~\bibnamefont{Oh}},
  \bibinfo{author}{\bibfnamefont{C.~M.} \bibnamefont{Ko}},
  \bibinfo{author}{\bibfnamefont{S.~H.} \bibnamefont{Lee}}, \bibnamefont{and}
  \bibinfo{author}{\bibfnamefont{S.}~\bibnamefont{Yasui}},
  \bibinfo{journal}{Phys.Rev.} \textbf{\bibinfo{volume}{C79}},
  \bibinfo{pages}{044905} (\bibinfo{year}{2009}).

\bibitem[{\citenamefont{Adamczyk et~al.}(2012)}]{Adamczyk2012}
\bibinfo{author}{\bibfnamefont{L.}~\bibnamefont{Adamczyk}} \bibnamefont{et~al.}
  (\bibinfo{collaboration}{STAR Collaboration}), \bibinfo{journal}{Phys.Rev.}
  \textbf{\bibinfo{volume}{D86}}, \bibinfo{pages}{072013}
  (\bibinfo{year}{2012}).

\end{thebibliography}

\end{document}